
\magnification=\magstep1 \baselineskip 20truept \def\eq{$$} 
\def\en{$$\medskip\noindent} \def\x{{\bf \nabla}} \def\B{\bf B} \def\gtsim
{>\kern-1.2em\lower1.1ex\hbox{$\sim$}} \def\ltsim
{<\kern-1.2em\lower1.1ex\hbox{$\sim$}} \def\b{\bf b} \def\v{\bf v} \def\k{\bf
k} \def\p{\parallel} \def\t{\perp} \def\vr{\langle |{\tilde v}_{\p}({\bf
k},T)|^2 \rangle} \def\ref{\par\noindent\hangindent 20pt}

\newbox\grsign \setbox\grsign=\hbox{$>$} \newdimen\grdimen
\grdimen=\ht\grsign \newbox\simlessbox \newbox\simgreatbox
\setbox\simgreatbox=\hbox{\raise.5ex\hbox{$>$}\llap
   {\lower.5ex\hbox{$\sim$}}}\ht1=\grdimen\dp1=0pt
\setbox\simlessbox=\hbox{\raise.5ex\hbox{$<$}\llap
   {\lower.5ex\hbox{$\sim$}}}\ht2=\grdimen\dp2=0pt
\def\gsim{\mathrel{\copy\simgreatbox}} \def\lsim{\mathrel{\copy\simlessbox}}

\centerline{\bf GENERATION OF DENSITY PERTURBATIONS BY}  \centerline{\bf
PRIMORDIAL MAGNETIC FIELDS}  \bigskip\bigskip \centerline{Eun-jin Kim,$^1$
Angela Olinto,$^2$ and Robert Rosner$^2$} \medskip \centerline{\it
$^1$Department of Physics} \centerline{\it $^2$Department of Astronomy \&
Astrophysics} \centerline{\it and} \centerline{\it Enrico Fermi Institute,
The University of Chicago} \centerline{\it 5640 S.\ Ellis Ave, Chicago, IL\ \
60637} \vskip 2.5cm \centerline{\bf Abstract} \medskip \noindent We study the
generation and evolution of density perturbations and peculiar velocities due
to primordial magnetic fields. We assume that a random magnetic field was
present before recombination and follow the field's effect on the baryon
fluid starting at recombination. We find that magnetic fields generate
growing density perturbations on length scales larger than the magnetic Jeans
length, $\lambda_B$, and damped oscillations for scales smaller than
$\lambda_B$.  For small  wavenumbers $k$ (large length scales), we find the
magnetic field-induced density power spectrum generally scales as $k^4$. We
derive the magnetic Jeans length explicitly by including the back--reaction
of the velocity field onto the magnetic field by decomposing the magnetic
field into a force-free background field and perturbations about it.
Depending on the strength of the magnetic field and the ultraviolet cutoff of
its spectrum, structure can be generated on small or intermediate scales
early in the history of the universe. For a present {\it rms} magnetic field
of  $10^{-10}$ G on intergalactic scales, we find that perturbations on
galactic scales could have gone non--linear at $z \simeq 6$. Finally, we
discuss how primordial magnetic fields affect scenarios of structure
formation with non--baryonic dark matter.

\vfill\eject

\bigskip \centerline{\bf 1.\ Introduction} \medskip

The past decade has seen tremendous growth in our observational picture of
the universe. The Cosmic Background Explorer (Smoot et al.\ 1992) and other
cosmic background experiments have shown that the large-scale clustering seen
in galaxy surveys is consistent with a primordial origin for density
perturbations. On the largest scales, where density perturbations are linear
(i.e., {\it rms} variations in the density are smaller than the mean),
microwave anisotropy observations point toward an approximately
Harrison-Zel'dovich spectrum of initial density perturbations ($P(k) \sim
\langle |\delta_k|^2 \rangle \propto k^n$, with $n \sim 1$; Smoot et al.\
1992; Ganga et al.\ 1993). Such a spectrum arises naturally in inflationary
theories as well as in models based on topological defects.

On scales smaller than $\sim 8 h^{-1}$ Mpc (where $h$ is the Hubble constant
in units of 100 km/sec/Mpc), galaxy clustering is non--linear, and one needs
to rely on numerical studies to get insight into the physics of cluster and
galaxy formation. (Hereafter, we set $h=1$ unless otherwise noted.) Not only do
the density perturbations become non--linear, but the complexity of the physics
involved escalates as hydrodynamical effects become important. In this paper,
we show that an element of this increased complexity that is often neglected,
namely magnetic fields, may play a key role in the formation of structure in the
non--linear regime. 

Interest in the possibility of primordial fields has also been rekindled by
recent studies of galactic dynamos, which suggest that the ``classical"
dynamo mechanism for galaxies, i.e., ``$\alpha-\omega$" mean field dynamos,
provide an inappropriate description of the origins of galactic magnetic
fields (Ko \& Parker 1989; Rosner \& DeLuca 1989; Kulsrud \& Anderson 1992;
Chakrabarti,  Rosner, \& Vainshtein 1994). A variety of difficulties have
been clearly recognized: To begin with, virtually all extant galactic dynamo
models are kinematic, and hence cannot properly describe the evolution of the
$\beta$ ($\equiv 8\pi p/B^2$, $p$ the gas pressure) $\approx 1$ galactic
magnetic field. Buoyancy is typically parametrized, and not explicitly taken
into account (despite the fact that it is probably the dominant process by
which magnetic fields generated in the solar interior and the galactic disk
interior reach the outer ``surfaces" of these objects; Parker 1979.).\ \ The
fields are assumed to be smooth; that is, there is usually no account taken
of the observed filamentation and other fine structure evident in both solar
magnetic fields and magnetic fields in our galaxy (although in the galactic
case, it is not certain that the structure is in the field rather than in,
for example, the particle density distribution). The galactic fluid is viewed
as a single-component gas; this means that the effects of neutrals (including
such phenomena as ambipolar diffusion) are ignored, contrary to what one
would expect (in contrast, cf.\ Kulsrud 1988; Zweibel 1988). Finally, there
is some considerable doubt as to the actual value of the turbulent (eddy)
diffusivity: It has been argued by Ko \& Parker (1989) that conditions
appropriate to magnetic field generation and turbulent diffusion occur rather
sporadically, and so are not the norm (also Parker 1992); and various authors
have recently argued that the concept of turbulent diffusion for magnetic
fields is itself flawed (cf.\ Vainshtein \& Rosner 1991; Kulsrud \& Anderson
1992; Cattaneo \& Vainshtein 1991; Cattaneo 1994; Gruzinov \& Diamond 1994).
Thus, although it is by no means established that dynamo models to explain
galactic fields cannot be constructed (indeed, see Parker 1992 for a recent
new proposal), it is by the same token no longer certain that extremely weak
primordial ``seed fields" are sufficient to account for the observed fields
in galaxies.

Of great relevance to understanding the role of magnetic fields in galaxy
formation are observations of intergalactic magnetic fields and magnetic
fields at high redshifts. Reports of Faraday rotation associated with
high-redshift Lyman-$\alpha$ absorption systems (Kronberg \& Perry 1982,
Wolfe 1988; Wolfe, Lanzetta, \& Oren 1991, Kronberg, Perry,   \& Zukowski,
1992) suggest that dynamically significant magnetic fields may be present in
condensations at high redshift. Together with observations of strong magnetic
fields in clusters (Kronberg 1994), these observations support the idea that
magnetic fields play a dynamical role in the evolution of structure.

The notion that magnetic fields may play an important role in structuring the
universe is not new. The most detailed study was done by  Wasserman (1978),
who assumed the existence of a random magnetic field at recombination and, by
treating it as a ``source term," showed that it may act as a source for
galaxy-scale density fluctuations. These calculations did not take into
account the possibility of fluid back-reactions to the Lorentz force
via the induction equation. Hence,
the temporal evolution of the magnetic fields was entirely due to the Hubble
expansion, and a magnetic Jeans length could not be derived. As we discuss
below, Wasserman's approach is valid for length scales much larger than the
magnetic Jeans length, $\lambda_B$,  and fluid velocities much smaller than
the Alfven speed, $v_A = B / \sqrt{ 4 \pi \rho}$. For velocities approaching
$v_A$, the back-reaction of the fluid onto the field cannot be neglected. We
study the $v \sim v_A$ regime by assuming that the background field is close
to force free; this treatment allows us to derive the magnetic Jeans length
within a linear analysis.

In order to describe the implications of random magnetic fields present at
recombination for structure formation, we consider the coupled evolution of
density perturbations, peculiar velocities, {\it and} magnetic fields for the
following two cases. First, we study the effect of a background magnetic
field to first order in the velocity and density fields (following Wasserman
1978) and deduce the spectrum of density perturbations generated on scales in
which the back-reaction onto the field is small, i.e., on scales   $\lambda
\gg \lambda_B$.  Second,  we include the fluid back--reactions to the magnetic
forces, and consider how this type of dynamics determines the power spectrum
of the resulting velocity field and density fluctuations when the background
magnetic field is close to force--free. Inclusion of the back-reaction onto
the magnetic field allows us to derive the magnetic Jeans length for this
problem, while the assumption of a force-free background field allows a linear
treatment of  scales $\lambda \sim \lambda_B$. Thus, we present a consistent
linear perturbation analysis of the combined magnetic--fluid evolution
equations (in the single--fluid approximation), and compute the present
density fluctuations  and vorticity under the assumption that random magnetic
fields existed before recombination. We show that the resulting spectrum for
density perturbations has a general form which is insensitive to the magnetic
field spectral index; on large scales the spectrum of density perturbations
is too steep ($P(k) \propto k^4$) to fit the observed spectrum, while on
small scales magnetic fields introduce a peak in the spectrum around  $k_{\rm
peak} \simeq {\rm min}(k_B, k_{\rm max})$, where $k_B= 2 \pi / \lambda_B$
and  $k_{\rm max}$ is the ultraviolet cutoff of the magnetic field spectrum.

The outline of our paper is as follows: We present the basic
magnetohydrodynamic equations used in our analysis, discuss plausible initial
conditions, and carry out our linearization, in \S 2. In \S 3, we discuss the
case studied by Wasserman and calculate the spectrum of density perturbations
for this case. In \S4, we consider the case of close to force-free background
magnetic fields and compute the time evolution and spectrum for the
compressible modes, the
 resulting density fluctuations, and the incompressible modes.  Our results
are discussed in \S 5. For the sake of clarity, we have placed details of our
analysis in the Appendices.

\bigskip\bigskip \centerline{\bf 2.\ The Perturbation Analysis} \medskip
 
In this section, we review the physical conditions at the time of
recombination, discuss the basic equations used in our analysis, and develop
our perturbation scheme.

We assume that random magnetic fields were present at the epoch of
recombination and that these were formed through pre-recombination processes
(e.g., Hogan 1983; Turner \& Widrow 1988; Quashnock, Loeb, \& Spergel 1989;
Vaschaspati 1991; Ratra 1992a,b; Cheng \& Olinto 1994). As the universe cooled
through recombination, baryons decoupled from the background radiation, and
the baryon Jeans length decreased from scales comparable to the Hubble scale
($\sim 100 $ Mpc) to $\sim 10 $ kpc in comoving units. (Throughout this
paper, we use $\lambda$ to describe comoving length scales and  set the
scale factor today to unity, $R(t_0) \equiv 1$; physical length scales, $l$, 
at any time can be found by multiplying the comoving scale by the
scale factor, $l(t)= \lambda R(t)$.)\ \ After recombination, baryons are free 
to condense on scales larger than the Jeans length, and will do so via
gravitational instabilities if there are initial perturbations in the density
field. Concurrently, magnetic fields that were frozen into the baryon-photon
plasma before recombination will tend to relax into less tangled configurations
once the baryons they are coupled to decouple from the photon background.
Consequently, density perturbations in the baryons can be generated through the
Lorentz force even if the density field is initially smooth, and the initial
peculiar velocity field vanishes at recombination. To understand the effect of
magnetic fields on the origin of density perturbations, we assume that no
initial density perturbations or peculiar velocities were present at
recombination, so that all subsequent density perturbations or velocities are
induced by magnetic fields alone. (We address the more general case of
combining initial density perturbations and magnetic field effects in a
subsequent paper.)

To follow the evolution of the density, peculiar velocity, and magnetic field
after recombination, we write the basic one-fluid magnetohydrodynamic (MHD)
equations in comoving coordinates,
	 
\eq\eqalignno{ \rho\left({\partial_t{\v}}+{\dot R\over R}{\v} 
+{{\v}\cdot{\bf \nabla v} \over R} \right) &=-{{\bf \nabla}p\over R}-\rho
{{\bf \nabla}\psi \over R}+{({\bf \nabla}\times{\bf B}) \times {\B}\over 4\pi
R},&(1)\cr {\partial_t\rho}+3{\dot R\over R}\rho 
+{{\bf\nabla}\cdot(\rho{\v})\over R} &=0,&(2)\cr {\bf\nabla}^2\psi\over R^2
&=4\pi G[\rho - \rho_b(t)],&(3)\cr {\bf\nabla}\cdot{\B} &=0,&(4)\cr
\partial_t(R^2{\B}) &={{\bf \nabla}\times ({\bf v} \times R^2{\B}) \over R}
&(5)\cr} \en (cf.\ Wasserman 1978), where $\psi$ is the gravitational
potential, $R$ is the scale factor, $\rho_b = \rho_b (t)$ is the uniform
background density, and all other symbols have their usual meaning (we set $c
= 1$ throughout the manuscript). We have neglected all viscous and diffusive
terms because the relevant Reynolds numbers are very large.

\medskip \centerline{\it 2.1 Initial Conditions and Background Evolution}
\smallskip

We begin with the basic assumption that all baryonic matter is uniformly
distributed at recombination ($t_{\rm rec}$), with density $\rho_b (t_{\rm
rec})$ and  that this matter has zero peculiar velocity, ${\v}({\bf x},t_{\rm
rec}) = 0$. Furthermore, we assume that there is a magnetic field already
present, ${\B} ({\bf x}, t_{\rm rec}) = {\B}_{\rm rec}({\bf x})$, presumably
created well before recombination; we posit that this magnetic field is randomly
oriented on spatial scales smaller than the Hubble radius at recombination {\it
and} has no ensemble averaged mean components on the Hubble scale; thus, we
assume that $ \langle {\B}_{\rm rec} ({\bf x}) \rangle = 0$, while $ \langle
|{\B}_{\rm rec} ({\bf x})|^2\rangle \neq 0$,  where the angular brackets mean
the ensemble average. 

The above magnetic field does not significantly perturb the baryonic matter
until photons and baryons decouple; once decoupling occurs, the unbalanced
Lorentz forces act to disturb the smooth background density,
${\rho}_b(t)$, leading to both density perturbations, $\delta \rho({\bf
x},t)$,  and peculiar velocities, ${\v}({\bf x},t)$, of the baryonic fluid
(Wasserman 1978). We decompose the fluid density, $\rho ({\bf x},t)$, and the
magnetic field, ${\B}({\bf x},t)$ by writing
 $$  \eqalignno{ \rho ({\bf x},t) &={\rho}_b(t)+\delta \rho({\bf x},t)
 \equiv \rho_b(t)[1+\delta({\bf x},t)] \,, \cr  {\B}({\bf x},t) &={\B}_b({\bf
x},t) +{\delta \B}({\bf x},t) \,. \cr }  $$   
Here $\B$ is the total field, ${\B}_b$ is the
background random magnetic field, with 
$$ {\B}_b({\bf x},t) = {\B}_{\rm rec}({\bf x}) {R^2(t_{\rm rec})\over R^2(t)}\,,
$$
 and ${\delta \B}$ is the
difference between the total field and the background; in other words,
${\B}_b (t)$ is simply the initial random field evolved only by the Hubble
flow, while $\delta \B$ is the additional field which results as the Lorentz
force perturbs the baryonic fluid, and the fluid reacts back. 

With these definitions, we assume
that at $t_{\rm rec}$: $$ \eqalignno{
 \delta({\bf x},t_{\rm rec}) 
 &=0 \,,\cr {\v} ({\bf x},t_{\rm rec}) 
 &=0 \,,\cr {\delta \B}({\bf x},t_{\rm rec})
 &=0 \,.\cr} $$ 

In addition, the temporal behavior of the background quantities is determined
by  the cosmological model we assume. To isolate the effect of magnetic
fields from other sources of density perturbation and to keep the analysis
simple, we chose to study first the case of a flat universe with a critical
density of baryons ($\Omega_{\rm baryon} = 1$). In  \S 3, \S 4, and \S 5,  we
discuss the effect on our results of including non-baryonic dark matter as
the dominant component of the universe.

For an Einstein--De Sitter model, the following relations hold during the
time of interest ($0\leq z \leq z_{\rm rec} \simeq 1100$): $$  \eqalignno{
\rho_b(t) R^3(t) &= \rho_b(t_0) \equiv \rho_0,&(6a)\cr 6\pi G \rho_b(t) t^2
&=1,&(6b)\cr R(t) &=(t/t_0)^{2/3},&(6c)\cr } $$ where $t_0$ is the age of the
universe, and quantities with subscript 0 are evaluated at the present time 
(we chose $R(t_0) = 1$). The background magnetic field satisfies 
$${\B}_b({\bf x},t)R^2(t) = {\B}_{\rm rec}({\bf x}) R^2(t_{\rm rec}) =
{\B}_b({\bf x},t_0) \equiv  {\B}_0({\bf x}) \,. \eqno(6d)$$

It is also useful to define  $$ \langle B_0^2 \rangle \equiv  \langle
|{\B}_0({\bf x})|^2 \rangle   = 8 \pi R(t)^4 \rho_B(t)  \ , \eqno(7) $$  where
$\rho_B(t)$ is the average magnetic field energy density present in the
universe at recombination redshifted to time $t$. 

In analogy with the ordinary Jeans length, $ \lambda_J = v_s \sqrt{\pi / \rho
G}$, one can expect the magnetic Jeans length to be  $$ \lambda_B \simeq v_A
\sqrt{\pi / \rho G}$$  (Peebles 1980). In \S 4, we show that the magnetic
Jeans length, defined by the wavenumber where the  transition  between
growing (or decaying) modes and oscillatory modes occurs, can be written as:
$$ \lambda_B \equiv {2 \pi \over k_B} = {2 \over 5 \rho_0} \sqrt{ \langle
B_0^2 \rangle \over G} \ . \eqno(8)$$

The magnetic Jeans length written above can be constrained by requiring that,
at the time of nucleosynthesis, the average energy density in the magnetic
field,  $$ \rho_B(t) = {\langle |{\B}({\bf x},t)|^2 \rangle \over {8 \pi}} =
{\langle B_0^2 \rangle \over {8 \pi R(t)^4}} \,, $$ be significantly less
than the energy density in radiation, $$ \rho_r(t)= { {\pi^2 g_* T^4} \over
30} \,, $$ where $g_*$ is the number of relativistic degrees of freedom, and
$T$ is the temperature of the background radiation (Barrow 1976, see also
Cheng, Schramm, \& Truran 1994,  Grasso \& Rubinstein 1995, and
Kernan, Starkman,  \& Vachaspati 1995). Since both energy densities redshift as
$R^{-4}$, it suffices to require that $\rho_B \ll \rho_r$ today, which gives:
$$ \langle B_0^2 \rangle^{1 \over 2} \ll 4 \times 10^{-6} ~ {\rm G} \,. 
$$ 
 By assuming a critical density in baryons, $\rho_0 = 2 \times 10^{-29}$
g cm$^{-3}$, in eq.\ (8), we can write: $$ k_B \simeq { {2 \pi }\over {100
 {\rm Mpc}}} \; \left({
 4 \times 10^{-6} {\rm G} \over {\langle B_0^2 \rangle^{1\over 2}}
}\right) \ \ . \eqno(9) $$ 
 Therefore, the constraint $\rho_B \ll \rho_r$ implies that the magnetic
Jeans length today must satisfy $ l_B(t_0) = \lambda_B \ll 100$ Mpc, i.e., 
$$ k_B \gg {{ 2 \pi }\over 100~ {\rm Mpc}} \ . \eqno(10) $$ (Note that
if we had set $\Omega_{\rm baryon} <1$ in the density used in eq.\ (8), the
effect would be to increase the maximum $\lambda_B$, thus weakening the
constraint.)

It is interesting to note that, unlike the ordinary Jeans length, $\lambda_B$
does not change with time; for $t_{\rm rec}< t <t_0$, $l_B (t) =
\lambda_B R(t)$. 

There are several observational constraints on a primordial magnetic field,
the  most stringent being an upper limit on the Faraday rotation of light from
distant quasars which requires that the average magnetic field on scales
comparable to the present horizon, $\bar B_0(H_0^{-1}) \lsim 10^{-11}$ G
(Kronberg \& Simard-Normandin 1976), where $H_0^{-1} \simeq 3$ Gpc.  Using
similar techniques to test the magnetic field on Mpc scales gives limits of  $
\bar B_0({\rm Mpc}) \lsim 10^{-9}$ G, which is consistent with the limit one
gets by assuming that the observed galactic magnetic field is solely due to a
primordial field enhanced by the collapse of the Galaxy. Expanding the volume
occupied by a galactic field of $\bar B_{gal} \simeq 3 \mu$G in order to
decrease the density from that in the Galaxy to the critical density gives an 
average field in the universe on the scale $l_g \sim$ Mpc (the
comoving scale that collapsed to form a galaxy) of $\bar B_0(l_g) \lsim
10^{-9}$ G (using $2 \times 10^{-24}$g cm$^{-3}$ for the average gas density in
the Galaxy, $2 \times 10^{-29} $g cm$^{-3}$ for the average gas density in the
universe, and the assumption that the field is frozen in as the gas contracts,
$B \propto \rho^{2/3}$). This argument is based on the simple assumption that
the magnetic field is frozen--in and grows with the collapse as $B \propto
\rho^{2/3}$.
 Kulsrud (1995) has  argued that by taking into account the formation and
rotation of the Galactic disk a seed field above $\bar B_0(l_g) \gsim
10^{-12}$ G may give rise to stronger galactic fields than observed. The
dynamics of the Galactic field once fields close to or even above
equipartition are reached has not been fully understood yet, so we will take
the limit on Mpc scale fields to be $\bar B_0({\rm Mpc}) \lsim 10^{-9}$ G
below.

 These limits cannot be unambiguously translated into a limit on $\langle
B_0^2 \rangle^{1 \over 2}$ since $\bar B_0(l)$ refers to the average field on
a particular scale $l$, for example,  averaged with a window function
(cf. eq.\ (42) below). This
averaging procedure depends not only on the integrated power spectrum
$\langle B_0^2 \rangle^{1 \over 2}$, but on the functional form of the power
spectrum ${\tilde B}^2(k)$ as well, unless $\langle B_0^2 \rangle^{1 \over 2}
\lsim 10^{-9}$ G. We  return to these constraints  during our discussion
of magnetic field spectra in \S 3 and \S 4.

\medskip \centerline{\it 2.2 The Perturbation Scheme} \smallskip
 
The next step is to identify our ``small" quantities, which will fix the
ordering of the perturbation scheme. In the spirit of a linearized theory, we
shall assume that the density perturbations resulting from the Lorentz force
are small, i.e., that $$ \delta \ll 1 \,. $$ Similarly, we assume that the
induced peculiar velocities and magnetic fields are small, e.g., we assume
that  $$ {{v_{\rm rms} \tau} \over {L}} \ll 1  \, , \  {\rm and} \ { |\delta
{\B}| \over |{\B}_b|} \ll 1 \, ,$$  where $\tau$ is the characteristic time
scale of the flow, $v_{\rm rms}$ is the {\it rms} flow speed, and $L$ is the
characteristic length scale of the flow at time $t_{\rm rec} + \tau$. Using
these scaling relationships, we linearize eqs.\ (1)--(5), noting that since
we are primarily interested in wavelengths larger than the ordinary Jeans
length, the pressure term may be ignored (Peebles 1980; 1993). 
Then, upon retaining leading order
terms, we obtain  $$ \eqalignno{ {\partial_t {\v}}+{\dot R\over R}\v  &=-{\x
\psi \over R} + {(\x\times {\B}_b ) \times {\B}_b \over 4 \pi R(t)\rho_b(t)},
&(11)\cr {\partial_t \delta}+{\x\cdot{\v}\over R} &=0,&(12)\cr \x^2\psi &=4\pi
R^2 G\rho_b\delta,&(13)\cr \x\cdot{\B}_b=\x\cdot\delta \B &=0,&(14)\cr
{\partial_t (R^2{\B}_b)} &=0,&(15)\cr \partial_t (R^2{\delta \B} ) &={{\x}
\times ({\v}\times R^2 {\B}_b)\over R} . &(16)\cr} $$
 Note that eqs.\ (15) and (16) allow us to clearly distinguish between the
background field ${\B}_b$ and the perturbed field $\delta \B$. 

The evolution of $\v$ due to the Lorentz force depends non-linearly on 
$\B$. If we
assume that the background field  at recombination is not force-free,
$(\x\times {\B}_b ) \times {\B}_b \neq 0$, the background field will generate
a velocity field with ${\partial_t {\v}} \propto (\x\times {\B}_b ) \times
{\B}_b$. The fluid motion will affect the background field through eq. (16),
but to higher order in perturbation theory.  The fluid back reaction can be
neglected to lowest order as long as $| \delta {\B} | \ll | {\B}_b |$. This
is the case studied by Wasserman (1978), who  derived the time evolution of
density perturbations, which we briefly review in \S 3.1.  In \S 3.2 we
derive the spectrum of density perturbations for this case. 

The effect of the fluid motions on the magnetic field cannot be neglected
when the fluid velocities become comparable to the Alfven speed, $v_A = B /
\sqrt{4 \pi \rho}$. This can be seen by studying the effect of one $k$-mode
of ${\B}_b$ in eqs. (11) and (16). Eq. (11) gives   $$ { v \over v_A} \sim 
{k t \over R} v_A $$  while eq. (16) implies $$ { \delta B \over B_b} \sim 
{k t \over R} v_A {v \over v_A} \sim \left( {v \over v_A} \right)^2 \ . $$ 
Therefore, $\delta B \ll B_b$ if $v \ll v_A$, and the back-reaction of the
fluid onto the field can be neglected for scales $\lambda \gg \lambda_B$. 

This first order description in which the background field is frozen-in (only
redshifts) is physically reasonable for scales $ \lambda_B \ll \lambda \lsim
\lambda_H$, in which a tangled background field ``recently'' entered the
horizon and had no chance to relax to a force-free configuration. For smaller
scales, $\lambda \sim \lambda_B$, an alternative description is necessary. At
recombination $\lambda_H$ is much larger than $\lambda_B$, for magnetic
fields of interest ($\langle B_0^2 \rangle^{1 \over 2}
 < 10^{-9}$ G implies $\lambda_B < 25 $ kpc), so scales close to
$\lambda_B$ have been inside the horizon for some time and will tend to relax
to force-free configurations.\footnote{$^1$}{There is a substantial
literature on the subject of how a magnetized plasma might relax in the
presence of constraints, starting with the classic papers of Woltjer (1958)
and Taylor (1974). For low-$\beta$ plasmas, Taylor argued that relaxation
occurs such that the total helicity, $H \equiv \int dr^3 ~ \vec A \cdot \vec
B$ (with $\vec A$ the vector potential corresponding to the magnetic field
$\vec B$) is constant; in that case, such a system will relax to its lowest
energy state consistent with this fixed value of $H$, with the important
property that the magnetic field is force--free, i.e., $\nabla \times \vec B
= \lambda (\psi) \vec B$, where $\psi$ is the flux function corresponding to
$\vec B$. Thus, as long as one can ignore the dynamical effects of the
baryons, one would expect the fields to relax to a force--free state.} At
recombination, intermediate scales ($\lambda_B \ll \lambda \ll \lambda_H$) will
be relaxing toward a force-free configuration while scales close to
$\lambda_B$ behave like oscillatory modes about a force-free configuration. 
For this range of scales, a more  appropriate description of the evolution of
the fluid and magnetic field can be constructed by assuming that the
background field has a  force-free component on intermediate to large scales,
${\B}_{ff}$, and small scale  perturbations about this  force-free
configuration, $\b$, $${\B}_b = {\B}_{ff} + {\b} \ . \eqno(17)$$
 By assuming that $|{\B}_{ff}| >> |\b|$, we can linearize eqs. (1) and (5)
by  keeping terms to first order in $\v$ and $\b$, and write
 $$ \eqalignno{ {\partial_t {\v}}+{\dot R\over R}{\v}  &=-{\x \psi \over R} +
{(\x\times {\B}_{ff} ) \times {\b} + (\x\times {\b} ) \times {\B}_{ff} \over
4 \pi R(t)\rho_b(t)} \ , &(18)\cr {\partial_t (R^2{\B}_{ff})} &=0 \ ,&(19)\cr
\partial_t (R^2{ \b} ) &={{\x} \times ({\v}\times R^2 {\B}_{ff})\over R} \ .
&(20)\cr} $$ 

Eqs. (18)--(20) together with eqs. (12)--(14) 
 allow us to study the behavior of the coupled equations when the fluid
back-reaction becomes significant and the magnetic Jeans length can be
derived.  This approach is valid when a separation of scales between the
force-free component and the perturbations about it is possible. To get the
spectrum of the generated density perturbations in this case, we need to
specify both the spectrum of ${\B}_{ff}$ and $\b$ at recombination
(${\B}_b({\bf x},t) R^2(t) / R^2_{\rm rec}= {\B}_{ff}({\bf x},t_{\rm rec}) + 
{\b}({\bf x},t_{\rm rec})$
and $ \delta {\B}({\bf x},t)= {\b}({\bf x},t) - {\b}({\bf x},t_{\rm rec})$).
Following this scheme, we  deduce the time evolution and  find $\lambda_B$  
in \S 4.1, and discuss the spectrum of the generated density perturbations in
\S 4.2.

The physical picture and plan of this paper is as follows: Starting at
recombination, we follow the effect of a random magnetic field present at
recombination on the velocity  and density fields. For scales $\lambda_B \ll
\lambda \lsim \lambda_H $, we use the linear approach developed by Wasserman,
who derived the time evolution of these fields. 
We recapitulate Wasserman's
results in \S 3.1 and derive in \S 3.2 the spectrum of generated density
perturbations by assuming either a power law or a delta function spectrum for
the magnetic field. We show that the power spectrum of density perturbations
is too steep to fit the observed clustering of galaxies on large scales. For
$\lambda \sim \lambda_B$, we derive in \S 4 the time evolution and spectra for
density and velocity fields by assuming a force-free background field and
either a power law or delta function spectrum for the magnetic field
perturbations about the force-free background.

\bigskip \centerline{\bf 3.\ Fixed Background Magnetic Field } \medskip

In this section, we first discuss the time evolution of the compressible
modes for the case in which the field has not relaxed to a force-free
background configuration and acts as a fixed comoving  source of  density
perturbations that only redshifts in time. This approach is valid to first
order in perturbation theory and for scales $\lambda \gg \lambda_B$. We then
solve for the spectrum of generated velocity field and density perturbations
and discuss the implications.

\medskip \centerline{\it 3.1 Time Evolution} \smallskip

When the back-reaction of the fluid onto the background magnetic field  may
be neglected,  (i.e., for $\lambda \gg \lambda_B$), we can solve for the time
evolution of $\delta({\bf x}, t)$ by taking the divergence of eq.\ (11) and
 using eqs. (12) and (13) (Wasserman 1978). This gives: $$ \partial_{tt}
\delta + 2 {\dot R\over R} \partial_{t} \delta -4 \pi G \rho_b \delta = 
{{\x} \cdot [(\x\times {\B}_b ) \times {\B}_b]  \over 4 \pi \rho_b R^2 } \ .
\eqno(21) $$ Making use of the time dependence of $R$, $\rho_b$, and
${\B}_b$, through eq. (6), we can write: $$ \partial_{tt} \delta +
{4 \over 3 t} \partial_{t} \delta -{2 \over 3 t^2} \delta =  \left( {t_0
\over t} \right)^2 {{\x} \cdot [ (\x\times {\B}_0 ) \times {\B}_0]  \over 4
\pi \rho_0} \ . \eqno(22) $$ (Note that in our convention $\rho_0 \equiv
\rho_b(t_0)$ while Wasserman uses $\rho_0(t)$ for the background density.)

The time evolution of $\delta$ is then given by  $$ \delta ({\bf x}, t) = 
t_0^2 {{\x} \cdot [(\x\times {\B}_0 ) \times  {\B}_0 ]  \over 4 \pi \rho_0}
\left[ {9\over 10} \left( {t \over t_{\rm rec}} \right)^{2 \over 3} + {3
\over 5 } { t_{\rm rec} \over t} - {3 \over 2} \right] \  . \eqno(23) $$ The
time evolution in this case is separable from the spatial dependence such
that every $k$-mode varies the same way in time; growing modes scale as
$t^{2/3}$ or as $R(t)$, while decaying modes scale as $t^{-1}$.

The time evolution of the compressible modes can be found via eq. (8), namely
$${\x} \cdot {\v} ({\bf x},t) =   { 3 t_{\rm rec} {\x} \cdot [(\x\times
{\B}_0 ) \times {\B}_0 ]  \over 20 \pi \rho_0} \left({t_0 \over t_{\rm rec}}
\right)^{4 \over 3} \left[  \left( {t \over t_{\rm rec}} \right)^{1 \over 3}
- \left({ t_{\rm rec} \over t} \right)^{4 \over 3} \right] \  . \eqno(24)$$
 
Finally, the time evolution of the incompressible mode is given by $$ {\x}
\times {\v} ({\bf x},t) =   { 3 t_{\rm rec} {\x} \times [(\x\times {\B}_0 ) 
\times {\B}_0]  \over 4 \pi \rho_0} \left({t_0 \over t_{\rm rec}} \right)^{4
\over 3} \left[  \left( { t_{\rm rec} \over t} \right)^{1 \over 3} - \left({
t_{\rm rec} \over t} \right)^{2 \over 3} \right] \  . \eqno(25)$$ Unlike
compressible modes, incompressible modes have no growing modes during the
matter dominated era.

\medskip  \centerline{\it 3.2 Spectra of Density Perturbations and Velocity
Fields } \smallskip

In order to obtain the spectral dependence of $\delta$ and ${\v}$ for a given
spectrum of magnetic fields at recombination, we define the
Fourier--transforms of the fluid variables in comoving coordinates, $$
\eqalignno{ {\delta}({\bf x},t) &=\int d^3{\k}~ \exp(i{{\k}\cdot {\bf
x}}){\tilde {\delta}}({\k},t) \,, &(26)\cr  {\v}({\bf x},t) &=\int d^3{\k}~
\exp(i{{\k}\cdot {\bf x}}){\tilde  {\v}}({\k},t) \,, &(27)\cr {\B}_0({\bf x})
&=\int d^3{\k}~ \exp(i{{\k}\cdot {\bf x}}){\tilde {\B}}({\k}) \,. &(28)\cr}
\en  ${\B}_{\rm rec} ({\bf x})$ is assumed to be homogeneous and isotropic
(and, obviously, so is ${\B}_0 ({\bf x})$); therefore $\tilde {\B} ({\k})$
obeys the relation (cf.\ Kraichnan \& Nagarajan 1967; Moffatt 1978) $$
\langle {\tilde {B}_i}({\bf k}_1)
 {\tilde {B}_j}^\ast ({\k}_2) \rangle  =
\delta^3({\k}_1-{\k}_2)\left(\delta_{ij}-{k_{1i}k_{2j}\over k_1^2}\right)
{{\tilde B}^2(k_1)\over 2} \,, \eqno(29) $$ where $i$ and $j$ label the
$i$-th and $j$-th components of the vector  ${\tilde {\B}}({\bf k})$. Note
that $$\langle B_0^2 \rangle \equiv 4 \pi \int dk k^2 
{\tilde B}^2(k)  = G \left( {5 \pi \rho_0 \over k_B}\right)^2  \ . \eqno(30)$$

For the case of a fixed magnetic background field considered in this section,
the spectral dependence of $\delta$ is solely due to the initial conditions at
recombination.  In \S 4, we discuss the case in which back-reactions become
important and the time dependence is a function of the wavenumber. Here, it
is sufficient to determine the spectral shape of $\partial_{tt} \delta$ at
recombination which is given by eqs. (12) and (24), 
$$  \partial_{tt} \delta |_{t_{\rm
rec}} = - {{\x \cdot {\partial_t {\v}} ({\bf x},t_{\rm rec})}\over  R_{\rm
rec}} = {\x \cdot[(\x\times {\B}_0)  \times {\B}_0] \over 4 \pi \rho_0 R_{\rm
rec}^3} \,. \eqno(31) $$ 

We would like to determine the spectral dependence of the ensemble average of
$\delta$ by assuming a spectrum ${\tilde B}^2(k) $ for a random  magnetic
field at recombination as in eq. (29). The ensemble average of both sides of
eq. (31) vanish, since there are as many positive and negative variations of
$\partial_{tt} \delta$, but the relevant quantities to be computed are the
ensemble averages of the  square of the fluid quantities. The root mean square
({\it rms}) of the Fourier--transform of the density field is defined by
 $$ {\tilde \delta}_{\rm rms} ({k},t) \equiv \sqrt { \langle |{\tilde
\delta}({\bf k},t)|^2  \rangle } . \ $$ Since, in  the fixed background field
case, the time  and spectral dependencies are separable, we can write  $$
{\tilde \delta}_{\rm rms} (k,t) \equiv c(k) \tau(t) \ , $$  where $$ \tau (t)
={9 \over 10} \biggl({t \over t_{\rm rec}}\biggr)^{2\over 3} + {3 \over 5}
{t_{\rm rec}\over t} - {3 \over 2}, \eqno(32) $$ 
 and    $c(k)$ is determined by eq.~ (31).  The spectral dependence of
${\tilde  \delta}_{\rm rms}$ is related to the  {\it rms} of the curl-free
component of the velocity field which we define by 
 $$  {\tilde v}_{\p  \rm rms}  ({k},t) \equiv \sqrt { \langle |{\tilde
v}_{\p}({\bf k},t)|^2 \rangle }  \ ,$$ where ${\k}\cdot {\tilde {\v}}({\k},t)
\equiv  k{\tilde v}_{\p}({\k},t)$, and the subscript ${\p}$ represents the
compressible (longitudinal) component of ${\tilde{\v}}$. Similarly, we define 
 $$  {\tilde v}_{{\t}  \rm rms}  ({ k},t) \equiv \sqrt { \langle |{\tilde
v}_{\t}({\bf k},t)|^2 \rangle }  \ ,$$ where ${\k} \times {\tilde
{\v}}({\k},t) \equiv  k{\tilde v}_{\t}({\k},t)$, and the subscript ${\t}$
represents the incompressible (transverse) component of ${\tilde{\v}}$.

The detailed derivation of $c(k)$ and ${\tilde v}_{\p {\rm rms}}(k,t)$ as a
function of ${\tilde B}^2(k)$ is provided in Appendix A; here we only write
the solution,   $$ c(k) = {{t_{\rm rec}}^2 \over R_{\rm rec}} \, {\dot
{\tilde v}}_{\p  {\rm rms}}(k,t_{\rm rec}) \, k \,, \eqno(33) $$ and
\eq\eqalignno{ {\dot {\tilde v}}_{\p  {\rm rms}}(k,t_{\rm rec})^2 \, k^2 = 
\left({1 \over 4 \pi \rho_0 R_{\rm rec}^2  }\right)^2 {V\over (4\pi)^2 } 
 & \int_{k_{\rm min}}^{k_{\rm max}}dk_1  \int^1_{-1}d{\mu} 
 {{\tilde B}^2(k_1){\tilde B}^2(|{\k}-{\k}_1|) \over |{\k}-{\k}_1|^2} &\cr &
&\cr &  \bigl \{2 k^5 k_1^3 \mu + k^4 k_1^4 ( 1- 5\mu^2) + 2 k^3 k_1^5 \mu^3
\bigr \}  \ , & (34)\cr}\en where $V$ is a volume factor and ${\bf k} \cdot
{\bf k}_1  = k k_1 \mu$.

 Since a given $k$-mode of the velocity field is a non-linear convolution of
the magnetic field spectrum, different $k$-modes of ${\tilde B}(k)$ will
contribute to a given mode of ${{\tilde v}(k)}$. In particular, if the
spectrum for the magnetic field has an ultraviolet cutoff, $k_{\rm max}$, the
velocity field modes will be non-zero for $k \le 2k_{\rm max}$.  In
principle, the lowest $k$-mode excited through eq.(31) is $k=0$, even if the
magnetic field spectrum has an infrared cutoff, but, in practice, the Hubble
radius at each time will provide an effective infrared cutoff for the density
perturbations excited by the magnetic field. As we show below, the power in
low $k$-modes is too small for the infrared cutoff to be relevant. 

In what follows, we discuss two Ans\"atze for the functional form of the
magnetic field spectrum: power laws and delta functions. The delta function
Ansatz is simple to calculate and can be used to relate our results to those
of Wasserman (1978) and to the more realistic case of the power law Ansatz. 

\smallskip \centerline{\it 3.2.1 Delta Function Spectra for $ {\tilde
B}^2(k)$ } \smallskip

If the magnetic field spectrum at recombination is a delta function, we can
write  $$ {\tilde B}^2(k) \equiv \langle B_0^2 \rangle {\delta (k-k_*) \over 
{4 \pi k^2}}\,. \eqno(35) $$ 
 The initial acceleration becomes for $k \le 2 k_*$,
 $$ {\dot{\tilde v}}_{\p  \rm rms}(k,t_{\rm rec}) = \sqrt{V} { \langle B_0^2
\rangle \over { (4 \pi)^3 \rho_0 R_{\rm rec}^2 }} \,  {k^{1 \over 2}\over
k_*} \eqno(36) $$ and zero otherwise (see Appendix A). Using eq.\ (33) and 
$k_B$ from eq.\ (30), we find for the  density spectrum $$ c(k) = {25
\sqrt{V} \over {384 \pi^2}} {k^{3 \over 2} \over {k_B^2 k_*}} \,. \eqno(37) $$

We can now calculate the power spectrum of density perturbations, $P(k,t)$,
and the variance ({\it rms} power per logarithmic wavenumber interval),
$\Delta (k,t)$, using the following definitions $$ P(k,t) \equiv {(2
\pi)^3\over V} \langle |{\tilde \delta(k,t)}|^2 \rangle  \eqno(38) $$ such
that, $$
 \langle |\delta ({\bf x},t)|^2 \rangle \equiv \int d^3{\k}P(k,t) \,,  $$
and  $$ \Delta(k,t)^2 \equiv 4 \pi k^3 P(k,t) = {32 \pi^4 \over V} k^3
\langle |{\tilde \delta(k,t)}|^2 \rangle \, \eqno(39) $$ such that, $$
 \langle |\delta ({\bf x},t)|^2 \rangle \equiv \int \Delta(k,t)^2 d{\rm ln} k
\, .  $$
For the delta function magnetic field spectrum, we obtain  $$ \Delta_{\delta}
(k,t) = {25 \sqrt{2} \over 96} \left({k \over k_B}\right)^3 {k_B \over k_*}
\, \tau (t) \, . \eqno(40) $$

The resulting power spectrum of magnetic field-generated density
perturbations can be compared to observations of galaxy clustering, if we
assume that galaxies trace the mass density distribution. The power spectrum
of  galaxy clustering has been measured over the range $0.1 ~ {\rm Mpc} \lsim
2 \pi k^{-1} \lsim 10^2$ Mpc (e.g., Geller \& Huchra 1989; Efstathiou
et al.\ 1990; Maddox et al.\ 1990; Collins, Nichol, \& Lumsden 1992; Fisher
et al.\ 1993), while information from anisotropies in the background
radiation reaches scales comparable to the present Hubble radius $2 \pi
k_0^{-1} \simeq
 3\times 10^3 $ Mpc (Smoot et al.\ 1992; Ganga, Cheng, Meyer, \& Page
1993). In the absence of bias, an estimate of the present-day scale of
non-linear clustering can be made by estimating the scale at which the {\it
rms} galaxy fluctuations are unity; for optically selected galaxies this is
$l_{nl}(t_0) = \lambda_{nl} \simeq 8$ Mpc. On large scales ($0.01 
{\rm Mpc}^{-1} < k < k_{nl}$), the observed galaxy power spectrum is
consistent with a power law, $P(k) \propto k^{n}$, $n \simeq -1$ with a hint
of a bend to larger $n$ at the larger scales. COBE suggests that $P(k)
\propto k$ on the largest scales. Therefore, magnetic field-induced
perturbations (with a delta function magnetic field spectrum) have too steep
a spectrum ($P_{\delta}(k) \propto k^3$) to agree with observations on large
scales. As we discuss below, a similar behavior is found if ${\tilde B}^2(k)$
is a power law (in that case, $P(k) \propto k^4$); therefore, magnetic
field-induced density perturbations {\it cannot} reproduce the observations
of structure on very large scales. 

On smaller scales ($k \gtsim \; 2 \pi /\lambda_{nl}$), magnetic field-induced
perturbations may play an important role. In particular, they are of interest
if ${\tilde B}^2(k)$ is such that $\Delta_{\delta}(k,t_0) \gtsim \; 1$ for
cosmologically relevant scales, say between clusters of galaxies ($k_{cl}
\simeq 2\pi/ 2 $ Mpc) and globular clusters ($k_{gl} \simeq 2\pi/
10^{-2} $ Mpc). As $\Delta_{\delta}(k,t)$ approaches 1, the linear treatment
used above  breaks down. However, we can make use of our linear solution to
approximately estimate the epoch, $t_{nl}(k)$, that a particular scale $k$
becomes non-linear by setting $\Delta (k,t_{nl}(k)) \simeq 1$.

Setting $\Delta_{\delta} (k,t_{nl}(k)) \simeq 1$, we find that $t_{nl}(k)$
satisfies
 $$ \left({t_{nl}(k) \over t_{\rm rec}}\right)^{2 \over 3}  = { {1 +z_{\rm
rec}} \over {1 + z_{nl}}} \simeq 3  {k_* \over k_B} \left({k_B\over
k}\right)^3 \,. \eqno(41) $$  The first scale to go non-linear in this case
is the smallest wavelength allowed or largest wavenumber $k= 2 k_*$. The time
at which perturbations with $k= 2 k_*$ become non-linear is then  $t_{nl}(2
k_*) \simeq 1.8 \; t_{\rm rec} \; ( k_B/ 2 k_*)^{3} $. Requiring that
$t_{nl}(2 k_*) \le t_0$ implies $2 k_* \gsim 3.7 \times 10^{-2} k_B$. For
example, suppose $2 k_* = 0.1 k_B$, which gives $t_{nl}(2 k_*)\simeq 1.8
\times 10^3 t_{\rm rec}$ or redshift $z_{nl}(2 k_*) \simeq 6$. In this case,
choosing the galaxy scale to go non-linear at $z \simeq 6$, we get $2 k_*
\simeq 2 \pi / $ Mpc, $k_B \simeq 2 \pi / 0.1 $ Mpc, and the
constraint eq.\ (10) is satisfied with $\langle B^2_0 \rangle^{1/2} \simeq 4
\times 10^{-9}$ G. This scenario would correspond to the formation of
galaxies around redshift 6. If instead, we choose cluster scales to go
non-linear at redshift $z_{nl} = 1$, then $2 k_* \simeq 2 \pi / 2$ Mpc
and $k_B \simeq 2 \pi / 0.1$ Mpc,  the same as the previous example. A
final example would be to choose the non-linear scale to be at present, 
$t_{nl}(2 k_*) = t_0$ and $2 k_* =2 \pi / 8 $ Mpc, then $k_B = 2 \pi /
0.3 $ Mpc, which corresponds to a field  $\langle B^2_0 \rangle^{1/2}
\simeq  10^{-8}$ G.

The examples given above satisfy the constraint that  $\rho_B \ll
\rho_{\gamma}$. As we mentioned before, there are other constraints on the
contribution of ${\tilde B}^2(k,t)$ to a given scale in which Faraday
rotation  measurements can be made. Unfortunately, these measurements can
only be translated to a limit on magnetic fields when assumptions  about the
density of ionized gas in the line of site are made. Since we have assumed a
critical density of baryons, applying the constraints from these measurements
can be somewhat ambiguous in the present discussion. When we address similar
effects in the presence of dark matter some of the ambiguity will not be
present (e.g., the dependence on $\Omega_{\rm baryon}$ can be kept explicitly).

Keeping in mind these limitations, we briefly discuss the implications of
these constraints for the examples given above. To apply these observational
constraints, we define the average field on scale $l$ by   \eq\eqalignno{
 \bar B^2 (l,t)  &\equiv \langle \left( \int d^3{\rm x} \; {\tilde{\B}}({\bf
x},t) W({\bf x}-{\bf x'}) \right)^2 \rangle_{\rm x'} \cr
 & = \int d^3k \;{\tilde B}^2(k,t) |W(kl)|^2 \ \ , &(42)  \cr} \en   where
$W({\bf x}-{\bf x}')$ is a window function which smoothes the magnetic field
on scale $l$ and $W(kl)$ is its Fourier transform.  Using a Gaussian window
function, $W({\bf x}-{\bf x}') = \exp (-|{\bf x}-{\bf x}'|^2/2 l^2)$ in eq.\
(36),  we find  $$ \bar B(l,t_0) = \langle B_0^2 \rangle^{1 \over 2} \exp
\left(- {k_*^2 l^2 \over 2 }\right) \,. $$  Requiring that $\bar B(l,t_0) \le
10^{-9}$ G for $l=l_g \simeq $ Mpc, we can write  $$ k_*^2 \ge {2
\over l_g^2} {\rm ln}\left( {2 \pi \over 25 {\rm kpc} }{1\over
k_B}\right) \,. \eqno(43) $$  For the three examples discussed above, the
formation of galaxies at $z=6$ and the formation of clusters at $z=1$ satisfy
eq.~ (43), while the parameters chosen for the formation of structure on scales 
$l_{nl}(t_0)$ today do not satisfy this constraint.

The examples discussed above demonstrate how non-linear structures within a
limited but relevant range of scales can be formed at reasonable redshifts
if  magnetic fields satisfying observational limits were present at
recombination. Because of the steep spectrum, the scales influenced by
magnetic fields are primarily in the non-linear regime, ultimately requiring
a detailed numerical study. Although the Ansatz used above for the magnetic
field spectrum is not realistic, some of the results obtained above are very
similar to the power-law spectrum discussed below.

Before leaving this section, we note that Wasserman (1978) discussed the case
in which ${\tilde B}^2(k)$ is sharply peaked around $k=2 \pi/ x_G$ (his $x_G$
corresponds to our $l_g$) and wrote (his eq.\ (26)) $$ \langle |\x
\cdot[(\x\times {\B}_{\rm rec}) \times{\B}_{\rm rec}]|^2 \rangle^{1\over 2}
\sim  \left({4\over 3}\right)^{1\over 2} \left({2\pi \over x_G}\right)^2
\langle B_{\rm rec}^2(x)\rangle \, $$ as an estimate for the effect of
magnetic fields. His result corresponds to an average over the volume $$
\langle |\x \cdot[(\x\times {\B}_{\rm rec}) \times{\B}_{\rm rec}]|^2 \rangle
= {( 4 \pi \rho_b(t_{\rm rec}) R_{\rm rec})^2 \over V} \int d^3 {\bf x}
\langle |{\x}\cdot {\dot {\v}}({\bf x},t)|^2 \rangle  $$  which, through
Parseval's theorem, can be re-written as $$ \langle |\x \cdot[(\x\times
{\B}_{\rm rec}) \times{\B}_{\rm rec}]|^2 \rangle = {( 4 \pi \rho_b(t_{\rm
rec}) R_{\rm rec})^2 \over V}  (2\pi)^3\int d^3 {\k} \langle |{\k}\cdot {\dot
{\tilde {\v}}}({\k},t)|^2 \rangle \,.  $$ Choosing $k_* = 2 \pi / x_G$ in the
delta function Ansatz for ${\tilde B}^2(k)$ and using eq.\ (36), we recover
Wasserman's eq.\ (26).

\smallskip \centerline{\it 3.2.2 Power Law Spectra for ${\tilde B}^2(k)$}
\smallskip

A variety  of mechanisms for generating magnetic fields before recombination
have been proposed (Hogan 1983; Turner \& Widrow 1988; Quashnock, Loeb, \&
Spergel 1989; Vaschaspati 1991; Ratra 1992a,b; Dolgov \& Silk 1993, Dolgov
1993; Cheng \& Olinto 1994), but consensus on a well--motivated  scenario is
still lacking. In general, most models generate a power law spectrum, ${\tilde
B}^2(k) \propto k^q$, with a cutoff on small scales (a typical cutoff scale is
some fraction of the Hubble radius when the field was generated). Of
particular interest is the case of a white noise ($q=0$) spectrum, which would
result from magnetic fields generated with similar strengths but random
directions within each Hubble volume during a phase transition in the early
universe (e.g., Hogan 1983). After the phase transition, this can be viewed as
a random walk of field lines, with stepsize of the order of the coherence
length (the Hubble radius at the phase transition, or some fraction of it). 

The magnetic field spectrum generated in the early universe will evolve
differently on different scales. On very large scales, the field is frozen
into the fluid and only redshifts with the expansion of the universe,
$B\propto R^{-2}$. On very small scales, the finite plasma conductivity
allows diffusion of the field within the plasma. On intermediate scales,
damping of the magnetohydrodynamic modes as the neutrinos decouple and later
as the photons decouple will change the effective cutoff  of the magnetic
field spectrum to scales as large as the Silk scale (Jedamzik, Katalinic, \&
Olinto 1995).

Here, we restrict our attention to the evolution starting at the end of
recombination for a magnetic field spectrum parametrized by the power law
index $q$ and an ultraviolet cutoff $k_{\rm max}$. We assume that, for
$k_{\rm min} \le k \le k_{\rm max}$,
 $$ {\tilde B}^2(k)\equiv A \; k^{q} \,, \eqno(44) $$  where $A$ is a
constant, $A \simeq (q + 3) \langle B_0^2 \rangle / k_{\rm max}^{q+3}$ \  (for
small $k_{\rm min}$). To  calculate ${\dot{\tilde v}}_{\p  \rm
rms}({k},t_{\rm rec})$ analytically, we make the further assumption that the
magnetic field is Gaussian distributed. The details of the tedious algebra
are left to Appendix A. To leading order in $k/k_{\rm max} \ll 1$ and for
integer spectral index $q$ between -1 and 6, we obtain the generic result
that   \eq\eqalignno{ {\dot{\tilde v}}_{\p \rm rms}({k},t_{\rm rec}) &\simeq
{\sqrt{V}} \; 2 \epsilon_q { \langle B^2_0 \rangle \over {(4 \pi)^3 \rho_0
R_{\rm rec}^2}} {k \over k_{\rm max}^{3 \over 2}}\,, &(45) \cr} \en  where  
$$ \epsilon_q \equiv {\sqrt{22} (q+3) \over 2 \sqrt{15 (2q+3)}} \,. $$  Using
eqs.\ (33) and (45), we find   
$$ c(k) = {25 \sqrt{V} \over {192 \pi^2}} {\epsilon_q k^
2 \over {k_B^2 k_{\rm max}^{3 \over 2}}} \ ,  $$  \smallskip $$ \Delta_{q}
(k,t) \simeq {25 \sqrt{2} \over 48} \epsilon_q \left({k \over k_B}\right)^{7
\over 2} \left({k_B \over k_{\rm max}}\right)^{3 \over 2} \, \tau (t) \ \ ,
\eqno(46) $$  and  $$ P_{q}(k,t) \simeq \left({25 \over 48}\right)^2
{\epsilon_q^2 \over {2 \pi k_{\rm max}^3}} \left({k \over k_B}\right)^4 \;
\tau (t)^2 \,. \eqno(47) $$

The spectrum of generated density perturbations is almost independent of the
magnetic field spectral index: $\epsilon_q \simeq 1$, for $-1 \le q \le 6$.
(This range in $q$ spans most of the proposed primordial magnetic field
spectra, with the exception of Ratra (1992a,b) who proposed indices up to $q=
-3$; we return to this issue in \S 4.2). 
In contrast, there is a strong dependence on
the ultraviolet cutoff $k_{\rm max}$, which plays a role similar to $k_*$ in
the delta function Ansatz. The amplitude of the spectrum of perturbations is
determined by $k_B$ and $k_{\rm max}$. For the power-law indices discussed
above, the dependence on an infrared cutoff, $k_{\rm min}$, is negligible
unless $k_{\rm min}$ is larger than the wavenumber of interest. 

The power spectrum is a power law, $P_{q} \propto k^4$, for scales inside the
horizon at recombination, i.e., for $k > k_{\rm rec} = 2 \pi H_{\rm rec}$,
and is steeper for $k < k_{\rm rec}$, with a   cutoff at $2 k_{\rm max}$. The
peak power is around  $k \sim 2 k_{\rm max}$ although the exact behavior
cannot be obtained from eq. (45), since we neglected terms of higher order in
$k / k_{\rm max}$.

An interesting property of the spectrum $P_{q}$ is that, unlike the case of
the delta function Ansatz, if we evaluate it at $k_B$, $P_{q}(k_B)$ does not
depend on $k_B$. This implies that if we extrapolate our solution to the
limit  $k \to k_B$,  $P_{q}$ would have  a fixed shape and amplitude as $k_B$
is changed, only shifting horizontally on a $P_{q}$ vs. $k$ plot.

Before exploring the relevant ranges in $k_{\rm max}$ and $k_B$ for the 
formation of structure,  we discuss the observational constraints on both
parameters. The constraint on $k_B$ in eq.\ (10) is unchanged, since it is
independent of the magnetic field power spectrum, while the constraint on
$k_{\rm max}$ depends on the spectral index $q$. For $q=0$, the average field
on scales $l$ is given by (from eqs.\ (42) and (44)),  $$ \bar B(l,t_0) =
\sqrt{ {3 \sqrt{\pi}\over 4}{ \langle B_0^2 \rangle \over k_{\rm max}^3 l^3}}
\ , \eqno(48) $$  where we use the same Gaussian window function as in the
previous section to smooth the field on  scales $l$. Again, for $l \simeq
l_g$, $ \bar B(l_g,t_0) \lsim 10^{-9}$ G and we get 
 $$ k_{\rm max} \gsim {1 \over l_g} \left( {2 \pi/25 {\rm kpc} \over
k_B}\right)^{2\over 3} \,, \eqno(49) $$ which is more restrictive than eq.\
(43). As $q$ increases, the constraint becomes less severe, since the steeper
the spectrum the less it contributes to $\bar{ B}_l$.
 
We now return to the ranges in $k_{\rm max}$ and $k_B$ that are relevant to
structure formation. In an analogous way to the delta function case, small
scales reach non-linear variance ($\Delta \ge 1$) earlier than large scales, 
and the first
scales to become non-linear have $k \sim 2 k_{\rm max}$. We again define
$t_{nl}(k)$, such that $\Delta_{q} (k,t_{nl}) =1$, which implies:   $$
\left({t_{nl}(k) \over t_{\rm rec}}\right)^{2 \over 3} \simeq {1.5 \over
\epsilon_q} \left({k_{\rm max} \over k_B}\right)^{3 \over 2 } \left({k_B\over
k}\right)^{7 \over 2} \,. \eqno(50) $$ 

Requiring that $t_{nl}(k \simeq 2 k_{\rm max}) \lsim t_0$ leads to the
constraint $ k_{\rm max}  \gsim 10^{-2} k_B$. This is an approximate estimate
for the lower limit for $k_{\rm max}$, since eq.\ (45) was derived in the
limit of $k \ll k_{\rm max}$.  If we again choose $ 2 k_{\rm max} = 2 \pi/ $
Mpc and require that this scale go non-linear at redshift $z_{nl} \simeq 6$, we
get $k_B = 2 \pi / 0.11 $ Mpc which satisfies both constraints from eq.
(10) and eq. (49).  This would correspond to the formation of galaxies at
redshift 6 for a background field $\langle B_0^2 \rangle \simeq 4.4 \times
10^{-9}$ G and a $\bar B(l_g)$ just below $10^{-9}$ G. If instead, we require
that magnetic fields form clusters of galaxies at redshift $z_{nl} = 1$, that
would correspond to $ 2 k_{\rm max} = 2 \pi/ 2$ Mpc and it would
require  $k_B = 2 \pi / 0.12$ Mpc. This scenario satisfies eq. (10)
easily and may or may not satisfy eq. (49) depending on the precise value for
the scale at which the constraint applies and  the value of the Hubble constant.
(For example, if we reintroduce the appropriate $h$-dependences, eq. (49) is
satisfied for $\bar B \le 10^{-9}$ G and  $h=0.5$ with $ l_g \ge
1.2 $ Mpc , or if $h=0.8$ then  $l_g \ge 1.7 $ Mpc.)  A scenario for generating
structure on scales which are going non-linear  today would correspond to 
choosing
the scale $2 k_{\rm max} = 2 \pi /8$ Mpc to go non-linear at $z_{nl} = 0$ which
satisfies eq. (10) but does not satisfy eq. (49). 
 
We see that with a more realistic choice for the power spectrum of magnetic
fields at recombination, non-linear structures in the cosmologically relevant
range of scales can be formed at reasonable redshifts for field strengths
which are below the present observational limits. Some of the examples above
have field strengths that are within the range of future observational
capabilities (Kronberg 1994; Lee, Olinto, \& Sigl 1995). 

We concentrated on galaxy and cluster scales in our examples above, but
another possible consequence of magnetic fields at recombination is the
formation of smaller objects, such as QSO's or Pop III stars, at very early
times. The challenge in the formation of smaller objects is the survival of
power on those scales through recombination, since photon drag damps 
magnetohydrodynamic
modes up to the Silk mass (Jedamzik, Katalinic, \& Olinto 1995). 

The scales on which magnetic fields generate structure are primarily in the
non-linear regime today, which limits our ability to make precise predictions
within linear perturbation theory. In the case of the baryonic universe
studied above, the range of scales affected by magnetic fields is quite
narrow, which suggests that objects formed via magnetic field perturbations
may be biased with respect to large scale structures formed by primordial
perturbations. 

For completeness, we derive the spectrum of generated incompressible modes.
The initial value of ${\dot{\tilde v}}_{\t {\rm rms}}(k,t_{\rm rec})$ can be
obtained from eq.\ (6) by taking the curl, $$ {\x } {\times}{\dot{\v}}({\bf
x},t_{\rm rec}) = {{\x\times[(\x\times{\B}_0)\times{\B}_0]}\over {4\pi \rho_0
R_{\rm rec}^2}} \, .  $$ Again, as in the compressible case $\langle
{\dot{\tilde v}}_{\t}({\k},t_{\rm rec}) \rangle =0$. 

If we assume a power law Ansatz for the magnetic field spectrum, as in eq.\
(44), we find for the incompressible velocity field spectrum: $$ {\dot{\tilde
v}}_{\t {\rm rms}}({k},t_{\rm rec}) \simeq {(q+3) \langle B^2_0 \rangle k
\over (4 \pi)^3 \rho_0
 R_{\rm rec}^2} {\sqrt{28 V\over 15(2q+3) k^3_{\rm max}}} \,.  $$ Comparing
to eq.\ (45), we see that the incompressible mode is initially excited to
almost the same extent as the compressible mode. Although the two modes have
comparable ``initial" amplitude and spectrum, they have quite different
temporal evolution and remain decoupled in the linear regime.

\bigskip \centerline{\bf 4.\ Force-Free Background Magnetic Fields} \medskip

In this section, we consider an alternative perturbation scheme more
appropriate for scales close to $\lambda_B$. We study the effect of magnetic
fields described by small scale perturbations about a large scale force-free
background field. We derive the time evolution of the compressible and
incompressible modes in \S 4.1 and in \S 4.2 we discuss the spectrum of the
generated modes for different magnetic field spectra.

\medskip \centerline{\it 4.1 Time Evolution} \smallskip

Taking the divergence and time derivative of eq.\ (18), and using  eqs.\
(12), (13), (19) and (20), we obtain an equation for the evolution of the
divergence of the velocity,   $$ {\partial_t}\left\{ R \, {\partial_t}
(R\x\cdot\v)\right\} - {4\pi G \rho_0 \over R} \x\cdot{\v} ={\x\cdot{\bf
Q}\over 4\pi\rho_0 R}\ , \eqno(51) $$  where   $${\bf Q} \equiv
[\x\times\x\times({\bf v}\times{\B}_{ff0})] \times {\B}_{ff0}
+(\x\times{\B}_{ff0})\times[\x\times({\bf v}\times{\B}_{ff0})]\ , \eqno(52) $$
and ${\B}_{ff0}$ is today's value for the force-free component of the magnetic
field. Since the perturbation scheme assumes $|{\B}_{ff0}| \gg |{\b}|$,
$|{\B}_{ff0}| \approx |{\B}_{0}|$. To leading order in this perturbation
scheme, the time evolution is determined by $|{\B}_{ff0}|^2$ with no 
contribution from ${\b}$, while the spectral dependence is linear in ${\b}$.
When deriving the time evolution and the magnetic Jeans length in this
section, we may use  the fact that $|{\B}_{ff0}| \approx |{\B}_{0}|$  to
omit the $ff$ label, but when deriving the spectrum in \S 4.2, 
it is important to keep the distinction. 

 Substituting the explicit time dependence given by eqs.\ (6) and (7), we can
re--write eq.\ (51) in the form   $$ \x\cdot \left(\partial_{tt}{\v} +{2\over
t} \partial_t{\v} -{4\over 9t^2}{\v}\right) = \left({t_o \over
t}\right)^2{\x\cdot{\bf Q}\over 4 \pi \rho_0 } \,. \eqno(53)  $$   The
evolution equation for each velocity mode ${\tilde {\v}} ({\k})$ is then
obtained by inserting Fourier expressions for the velocity field (as in eq.
(27)), and for the force-free background field in analogy to eq. (28). 
Calling ${\tilde {\B}}_{ff}({\k})$ the Fourier transform of ${\B}_{ff0} ({\bf
x}, t)$ as in eq. (28), we get:  \eq\eqalignno{ {\bf k} \cdot \left(t^2
\partial _{tt} + 2 t \partial _t -{4\over9}\right) & \ {\tilde {\v}}({\k},t)
= &(54)\cr 
  & \beta \int d^3 {\bf k}_1 d^3 {\bf k}_2 \ F[ {\bf k},{\bf k}_1, {\bf
k}_2,  {\tilde {\B}}_{ff} ({{\bf k}}_1),{\tilde{\B}}_{ff}({{\bf k}}_2),
{\tilde {\v}}({\k}-{\k}_1-{\k}_2,t)] , &\cr} \en  where

\eq\eqalign{ \beta &\equiv { t^2_0 \over 4 \pi \rho_0} = {1 \over 24 \pi^2
\rho_0^2 G} \,,\cr  F &\equiv -\{ k^2 (  {\tilde {\B}}_1 \cdot  {\tilde
{\B}}_2) (k-k_2)_i - k^2[({\k}-{\k}_2) \cdot   {\tilde {\B}}_1] { \tilde
B}_{2i}\cr &\;\; + 2({\k} \cdot  {\tilde {\B}}_2)[({\k}\cdot   {\tilde
{\B}}_1) k_{2i} - ({\k}_2 \cdot  {\tilde {\B}}_1)k_i]\} \, {\tilde
v}_i({\k}-{\k}_1-{\k}_2,t) \,,\cr} \en
 $  {\tilde {\B}}_1 \equiv   {\tilde {\B}}_{ff}({\k}_1)$, and $  {\tilde
{\B}}_2 \equiv {\tilde{\B}}_{ff}({\k}_2)$ . For notational convenience, we
define a logarithmic time variable $T \equiv \ln t$, and further define the
operator $G_i$ such that  $$ \int d^3 {\bf k}_1 d^3 {\bf k}_2 \ F \equiv G_i(
{\bf k:l}) v_i({\bf l}, T) \,, $$ i.e.,  $$ \eqalign{ G_i &\equiv -\int
d^3{\k}_1 d^3{\k}_2 d^3{\bf l} \{ k^2 (  {\tilde {\B}}_1 \cdot   {\tilde
{\B}}_2) (k-k_2)_i - k^2[({\k}-{\k}_2) \cdot   {\tilde {\B}}_1] {\tilde
{B}}_{2i}\cr &\;\; + 2({\k} \cdot   {\tilde {\B}}_2)[({\k}\cdot {\tilde
{\B}}_1) k_{2i} - ({\k}_2 \cdot   {\tilde {\B}}_1)k_i]\} \, \delta
({\k}-{\k}_1-{\k}_2-{\bf l}) \,.\cr} $$  Upon using this definition, eq.\
(19) then reads $$ \left(\partial _{TT} + \partial _T - {4\over 9}\right)\
{\tilde v}_{\p}({\k},T) ={\beta \over k}G_i( {\bf k:l}) {\tilde v}_i({\bf l},
T) \,. \eqno(55) $$ Since the ensemble average of both sides of eq.\ (55)
vanish, we compute the evolution of the quadratic quantities. At the expense
of yet further algebra, one obtains the following three equations:

$$ \eqalignno{ {\partial_{T}} \langle |{\tilde v}_{\p}({\k},T)|^2 \rangle  &=
2 Re \langle {\tilde v}_{\p}({\k},T){\partial_T {\tilde v}}_{\p}^*({\k},T)
\rangle \,, &(56a)\cr {\partial_T} Re \langle {\tilde
v}_{\p}({\k},T){\partial_T {\tilde v}}_{\p}^*({\k},T) \rangle &={4\over 9}
\langle |{\tilde v}_{\p}({\k},T)|^2 \rangle -
 Re \langle {\tilde v}_{\p}({\k},T){\partial_T {\tilde v}}_{\p}^*({\k},T)
\rangle &\cr + \langle | {\partial_T {\tilde v}}_{\p}({\k},T)|^2 \rangle  &+
{ \beta \over k} Re \langle G_i ({\k}:{\bf l}) {\tilde v}_i({\bf l},T)
{\tilde v}_{\p}^*({\k},T) \rangle \,, &(56b)\cr {\partial_T} \langle |
{\partial_T {\tilde v}}_{\p}({\k},T)|^2 \rangle  &= {8\over 9} Re \langle
{\tilde v}_{\p}({\k},T){\partial_T {\tilde v}}_{\p}^*({\k},T) \rangle -{2}
\langle |{\partial_T {\tilde v}}_{\p}({\k},T)|^2 \rangle &\cr &+{ 2\beta
\over k} Re \langle G_i ({\k}:{\bf l}) {\tilde v}_i({\bf l},T) {\partial_T
{\tilde  v}}_{\p}^*({\k},T) \rangle \,, &(56c)\cr} \en  where {\it Re} stands
for the real part. These equations look more complex than they really are
because the recurring term $\langle G v v^* \rangle $ can be calculated
easily if one notes that $ \langle B {\tilde v}(T) \rangle $ and $\int G
\langle v_{\p} v_{\t} \rangle$ vanish (see Appendix B). Thus,

$$
 \langle G_i ({\k}:{\bf l}) {\tilde v}_i({\bf l}) {\tilde v}_{\p}^*({\k})
\rangle  =-{2 \over 3} \langle B_{ff0}^2 \rangle k^3 \langle |{\tilde
v}_{\p}({\k})|^2 \rangle \,, $$ and $$ \langle G_i ({\k}:{\bf l}) {\tilde
v}_i({\bf l}) {\partial_T {\tilde v}}_{\p}^*({\k}) \rangle =-{2 \over 3}
\langle B_{ff0}^2 \rangle k^3 
 \langle {\tilde v}_{\p}({\k}) {\partial_T {\tilde v}}_{\p}^*({\k}) \rangle 
\,, $$ where $$ \langle B_{ff0}^2 \rangle \equiv  4 \pi \int dk \ k^2 
{\tilde  B}^2_{ff}(k) \ .$$ 

\noindent These considerations allow us to write a single evolution equation
for $\vr$, $$ \biggl\{{\partial_{TTT}} + 3 {\partial_{TT}} +\biggl({2 \over
9}+{4 \over 3} ak^2\biggr){\partial_T} +\biggl({4 \over 3} ak^2-{16 \over
9}\biggr)\biggr\} \vr = 0 \,, \eqno(57) $$ where $$ a \equiv 2 \beta \langle
B_{ff0}^2 \rangle = {\langle B_{ff0}^2 \rangle \over {12 \pi^2 \rho_0^2 G}}
\,. $$  In addition, we have the initial conditions (at $T=T_{\rm rec}$):  $$
\eqalign{
 \langle {\tilde v}_{\p}({\k},T=T_{\rm rec}) \rangle  &=0 \,,\cr {\partial_T}
\vr |_{_{T_{\rm rec}}} &=2 Re \langle {\tilde v}_{\p}({\k},T_{\rm
rec}){\partial_T {\tilde v}}^*_{\p}({\k},T_{\rm rec}) \rangle =0 \,,\cr
{\partial_{TT}} \vr |_{_{T=T_{\rm rec}}} &= 2 \langle | {\partial_T {\tilde
v}}_{\p}({\k},T_{\rm rec})|^2 \rangle = 2 \, [ t_{\rm rec} \, {\dot{\tilde
v}}_{\p  {\rm rms}} (k,t_{\rm rec}) ]^2 \,, \cr} $$ where ${\dot v} \equiv
\partial_t v$ and ${\tilde v}_{\p  {\rm rms}}(k,t) \equiv \sqrt { \langle
|{\tilde v}_{\p}({\bf k},t)|^2 \rangle }$. These follow from the assumption
that the velocity is zero  everywhere at $t=t_{\rm rec}$. In the last initial
condition above, $\dot {\tilde v}_{\p  {\rm rms}}(k,t_{\rm rec})$ will be
obtained later upon using eq.\ (18).

Returning to our original time variable $t\equiv \exp (T) $, and again
defining time derivatives with respect to the ordinary time $t$, the solution
for the {\it rms} velocity can be written as follows:
 
\medskip\noindent (i) For $k \ne k_B:$

$$ {\tilde v}_{\p  \rm rms}(k,t) ={\dot {\tilde v}}_{\p  \rm rms}(k, t_{\rm
rec}){{3 t_{\rm rec}}\over m} \biggl({t \over t_{\rm rec}}\biggr)^{-{1\over
2}}\biggl[\biggl({t \over t_{\rm rec}}\biggr)^{{m}\over 6}-\biggl({t\over
t_{\rm rec}}\biggr)^{-{m \over 6} }\biggr], \eqno(58a) $$ \medskip\noindent
(ii) For $k = k_B:$

$$ {\tilde v}_{\p  \rm rms}(k,t) ={\dot {\tilde v}}_{\p  \rm rms}(k,t_{\rm
rec})t_{\rm rec}\biggl({t \over t_{\rm rec}}\biggr)^{-{1 \over 2}}
\ln\biggl({t\over t_{\rm rec}}\biggr), \eqno(58b) $$ where $\ m(k) \equiv
{\sqrt{25-12ak^2}} = {5}{\sqrt{1-(k/k_B)^2}}$. The transition from stable
($m^2 < 0$) to unstable ($m^2 > 0$) modes occurs at $k=k_B$, and we therefore
find the magnetic Jeans length to be 

$$ \lambda_B \equiv {2 \pi\over k_B} = 2 \pi \sqrt{12a \over 25} = {2\over5}{
\langle B_{ff0}^2\rangle^{1/2} \over \rho_0 \sqrt{G}}\ . \eqno(59) $$  Modes
with length scales greater than $\lambda_B$ ($k < k_B$) are unstable, while
those on scales smaller than $\lambda_B$ ($k > k_B$) undergo damped
oscillations. As in Peebles (1980, 1993), one can roughly estimate the
magnetic Jeans length by using the Alfv\'en speed, $v_A = B / \sqrt{4 \pi
\rho}$, in place of the sound speed in the expression for the ordinary Jeans
length, which gives $$ \lambda_B \sim {B\over 2 \rho}{1\over \sqrt{G}} \,, $$
very close to the  expression derived in eq.\ (59).

We can now solve for $$ {\tilde \delta}_{\rm rms} (k,t) \equiv \sqrt {
\langle |{\tilde \delta}({\bf k},t)|^2 \rangle } $$  by using the
corresponding solution ${\tilde v}_{\p {\rm rms}}(k,t)$, with the initial
condition  $$ {\tilde \delta}_{\rm rms} (k,t_{\rm rec}) =0 \,, $$ and eq.\
(12). The detailed derivation of ${\tilde \delta}_{\rm rms} (k,t)$ as a
function of ${\tilde v}_{\rm rms}(k,t)$ is provided in Appendix C with the
result \eq {\partial \over \partial t} \, {\tilde \delta}_{\rm rms} (k,t) = k
{{\tilde v}_{\rm rms} (k,t)\over R(t)} \,. \en Thus, we can write the
solution,  $$ {\tilde \delta}_{\rm rms} (k,t) \equiv c(k) \tau(k,t) $$ where,
for $k \ne k_B$, $$ \tau (k,t) ={{18}\over m^2-1}\biggl[\biggl[1+{1 \over
m}\biggr] \biggl({t \over t_{\rm rec}}\biggr)^{{m-1}\over 6}+
\biggl[1-{1\over m}\biggr] \biggl({t\over t_{\rm rec}}\biggr)^{-{(m+1)\over
6}}- 2\biggr], \eqno(60a) $$ for $k = k_B$, $$ \tau (k,t) =36 \biggl({t \over
t_{\rm rec}}\biggr)^{-{1 \over 6}}\biggl[\biggl({t\over t_{\rm rec}}
\biggr)^{1 \over 6} - 1 - {1 \over 6}\ln\biggl({t\over t_{\rm
rec}}\biggr)\biggr], \eqno(60b) $$  and, again, $$ c(k) = {{t_{\rm rec}}^2
\over R_{\rm rec}} \, {\dot {\tilde v}}_{\p  \rm rms}(k,t_{\rm rec}) \, k \,.
\eqno(61) $$

In the form for ${\tilde \delta}_{\rm rms} (k,t)$ defined above, $\tau(k,t)$
can be thought of as a transfer function which evolves an initial spectrum
$c(k)$ from recombination to a later time $t$. We now discuss the behavior of
$\tau(k,t)$ and leave a discussion of $c(k)$, which depends on the particular
form of the magnetic field spectrum, to the next section (\S 4.2). 

An interesting feature of the solution is that $\tau(k,t)$ is independent of
the spectrum of the magnetic field (as long as its statistics are 
Gaussian)
and only depends on the combinations $ k / k_B$ and $t/t_{\rm rec}$.
Therefore, the solution can be easily rescaled for different choices of
magnetic field strength. The time evolution reduces to the usual $t^{2/3}$
growing and $t^{-1}$ decaying modes in the limit $k / k_B \to 0$. In this
limit,  $m \simeq 5$ and  the growth of perturbations is
nearly independent of $k$, so that the final spectrum is mostly proportional
to the initial spectrum given by $c(k)$ as in \S 3. As $k \to k_B$ from below,
the growth of perturbations decreases and, at $k = k_B$, the solution changes
from unstable (for $k < k_B$) to damped oscillatory (for $k > k_B$). 

In principle, the time evolution derived above is only valid for modes which
correspond to scales smaller than the Hubble radius at recombination, $k >
k_{\rm rec} \simeq 2 \pi / 100$ Mpc. Modes on comoving scales between the
Hubble radius at recombination and the Hubble radius today have similar but
delayed time evolution, since these modes start growing after they enter the
Hubble radius. As we show below, the spectrum of perturbations, $c(k)$, is
fairly steep for small $k$, so that the power for small $k$ modes is
negligible, and we can approximate the power for $k < k_{\rm rec}$ to be
zero. Alternatively, we can estimate the effect of the delayed evolution by
using $t/t_{\rm enter}(k)$ instead of $t/t_{\rm rec}$ in $\tau(k,t)$ for
modes with $k<k_{\rm rec}$, where $t_{\rm enter}(k)$ is the time a $k$-mode
enters the Hubble radius, $t_{\rm enter}(k) = t_{\rm rec} (k_{\rm rec} / k
)^3$. 

In Fig.\ 1, we plot $\tau (k,t/t_{\rm rec})$ for $k > k_{\rm rec}$ and $\tau
(k, t/t_{\rm enter})$ for $k<k_{\rm rec}$ at different redshifts. (The sharp
discontinuity at $k_{\rm rec}$ is an artifact of the approximation that
recombination happened instantaneously.) We can see that, independent of the
spectrum of the magnetic field, the power on scales $k \le k_{\rm rec}$ is
suppressed (by the delayed growth) as is the power on scales $k \ge k_B$.
Therefore, while magnetic fields do not generate significant clustering on
scales larger than $\sim 2 \pi k_{\rm rec}^{-1}$ and smaller than $\sim 2 \pi
k_B^{-1}$, they can have significant influence on the formation of structure
between these two scales, depending on the strength and spectrum of magnetic
fields at recombination.

If the universe were baryon--dominated with infinite conductivity, magnetic
fields could deter the growth of perturbations on scales smaller than
$\lambda_B$; hence, a constraint on the strength of magnetic fields could be
derived by the observation that structures do form on a given scale above the
ordinary Jeans length $\lambda_J$. For example, for galaxies to form in a
baryon--dominated, infinitely conducting flat universe, we would require
$\lambda_B \lsim l_{g} \simeq 1$ Mpc and, therefore, $\langle B_0^2
\rangle^{1/2} \lsim 4 \times 10^{-8}$ G. However, this result does not
take into account the presence of neutral hydrogen or non-baryonic dark
matter. When these components are present, modes on scales below $\lambda_B$
are also unstable. 

Finally,  the time evolution of incompressible modes can be derived in an
analogous manner to the derivation of the compressible mode evolution.  We
should also  note that the incompressible modes do not affect the
compressible modes just studied, and vice-versa, an assertion which can be
demonstrated by considering eq.\ (20) in detail.

In order to study the incompressible modes, we take the curl and the time
derivative of eq.\ (18), use eqs.\ (19)--(20), and find   $$ \x \times
\left(\partial_{tt}{\v} +{2\over t} \partial_t{\v} +{2\over 9t^2}{\v}\right)
= \left({t_0 \over t}\right)^2{\x \times {\bf Q}\over 4 \pi \rho_0 } \,, 
\eqno(62) $$ which is the incompressible analog of eq.\ (53). We then take
Fourier transforms as in eqs.\ (27)--(28), and obtain the analog of eq.\ (57)
for the incompressible mode:

$$ \biggl\{{\partial_{TTT}} + 3 {\partial_{TT}} +\biggl({26 \over 9}+{2 \over
3} ak^2\biggr){\partial_T} +\biggl({2 \over 3} ak^2+{8 \over
9}\biggr)\biggr\} \langle |{\tilde v}_{\t}({\k},t)|^2 \rangle = 0 \,,
\eqno(63) $$ where $ k{\dot{\tilde v}}_{\t}({\k},t_{\rm rec}) \equiv | {\k}
\times {\dot{\tilde {\v}}}({\k},t_{\rm rec})|$.

Since the calculation is essentially the same as that described above for the
compressible case, we only present the solution for the incompressible case
here: \medskip\noindent (i) for $ak^2 \ne 1/6$: $$ {\tilde v}_{\t {\rm
rms}}(k,t) ={\dot{\tilde v}}_{\t {\rm rms}}(k,t_{\rm rec}){t_{\rm rec}\over
2p} \biggl({t \over t_{\rm rec}}\biggr)^{-{1 \over 2}} \biggl[\biggl({t \over
t_{\rm rec}}\biggr)^{p}-\biggl({t\over t_{\rm rec}}\biggr)^{-p}\biggr] ,
\eqno(64a) $$ \smallskip\noindent (ii) for $ak^2=1/6$: $$ {\tilde v}_{\t
{\rm rms}}(k,t) ={\dot{\tilde v}}_{\t {\rm rms}}(k,t_{\rm rec})t_{\rm
rec}\biggl({t\over t_{\rm rec}}\biggr)^{-{1 \over 2}} \ln\biggl({t\over
t_{\rm rec}}\biggr) \,, \eqno(64b) $$ where ${\tilde v}_{\t {\rm rms}}(k,t)
\equiv \sqrt { \langle  |{\tilde v}_{\t}({\k},t)|^2 \rangle }$, $p(k)\equiv
{\sqrt{1-6ak^2}}/6$, and 
$a = {\langle B_{ff0}^2 \rangle /{12 \pi^2 \rho_0^2 G}}$
is the same as in the compressible case. Unlike the compressible modes, the
incompressible modes have no growing solution, only decaying or damped
oscillatory solutions.

\medskip \centerline{\it 4.2 The Spectrum of Compressible Modes} \smallskip

In order to obtain the spectral dependence of the generated compressible
modes, we need to find  ${\dot{\tilde v}}_{\p \rm rms}(k,t_{\rm rec})$ or
$c(k)$ for a given magnetic field spectrum at recombination.  From eq. (18):
 $$ \eqalignno{{\x \cdot {\dot{\v}}({\bf x},t_{\rm rec})}   &= {\x
\cdot \{ [(\x\times {\B}_{ff\rm rec}) \times{\b}_{\rm rec}] + [(\x\times
{\b}_{\rm rec}) \times {\B}_{ff\rm rec} ]\} \over 4 \pi \rho_b(t_{\rm
rec})R_{\rm rec}} &\cr  &= {\x \cdot \{ [(\x\times {\B}_{ff0}) \times{\b}_0]
+[(\x\times {\b}_0)  \times {\B}_{ff0}] \}  \over 4 \pi \rho_0 R_{\rm rec}^2}
\,,   &(65)\cr } \en and we need to specify spectral dependencies of both
$
{\B}_{ff0}$ and ${\b}_0$. 

In what follows, we discuss two Ans\"atze for the functional form of the
magnetic field spectrum. We study the solutions for the cases in which the
perturbations about the force-free field are either a delta function or  
power laws.   In both cases, we choose a delta function for the background
force-free spectrum to make the problem tractable analytically. This choice
adequately describes the force-free background as long as the scale of
perturbations onto this background are much smaller than the background
scale.

Assuming the force-free background magnetic field spectrum  has a spectrum
peaked at some scale, $\lambda_0 = 2 \pi / k_0$, we can describe it  by 
   $$ {\tilde B}_{ff}^2(k) \equiv \langle B_{ff0}^2 \rangle {\delta (k-k_0)
\over {4 \pi k^2}}\,. \eqno(66) $$ 

In Appendix D, we show that for this choice of $ {\tilde B}_{ff}^2(k)$ the
initial acceleration can be written as: \eq\eqalignno{ \langle |{\k}\cdot
{\dot {\tilde {\v}}}({k},t_{\rm rec})|^2 \rangle & =  {\alpha ^2 \langle
B_{ff0}^2 \rangle \,V  \over 4 (4\pi)^3\,k k_0^3} \int_{k_{\rm min}}^{k_{\rm
max}}\,dk_1{{\tilde b}^2(k_1) \over k_1}\, \biggl\{ k_1^8 -(4 k_0^2+2k^2)
k_1^6 +(6k_0^4+2k^2 k_0^2) k_1^4  &\cr & +(-4k_0^6+2 k^2 k_0^4 +16 k^4
k_0^2+2k^6)k_1^2 + k_0^8 -2k^2 k_0^6 +2k^6 k_0^2 -k^8 \biggr\}\,, &(67)\cr}
\en and ${\tilde b}^2(k_1)$ is specified in the next two sections.
 
\smallskip \centerline{\it 4.2.1 Delta Function Spectra for  $ {\tilde
b}^2(k)$ } \smallskip

If we assume that the deviations from the force-free background are also
peaked around a scale $k_b$, then we can write $$ {\tilde b}^2(k) \equiv
\langle b_0^2 \rangle {\delta (k-k_b) \over {4 \pi k^2}}\,, \eqno(68) $$
where  $$\langle b_0^2 \rangle \equiv 4 \pi \int dk k^2 {\tilde b}^2(k) \ .$$ 
The
spectrum of generated compressible modes is then given by 
 $$ {\dot{\tilde v}}_{\p  \rm rms}(k,t_{\rm rec}) =  {  \sqrt{V \langle
B_{ff0}^2 \rangle \langle b_0^2 \rangle} \over { (4 \pi)^3 \rho_0 R_{\rm
rec}^2 }} \,  {2 k^{1 \over 2}\over \sqrt{k_b k_0}} \sqrt{F_{\delta}(k, k_0,
k_b)} \eqno(69) $$ for $k_b - k_0 \le k \le k_b+k_0$, where $F_{\delta}(k,
k_0, k_b)$ is given in Appendix D, eq. (D7). Using eqs.\ (59) and (61), we
find for the ``initial" density spectrum  $$ c(k) = {25 \sqrt{V} \over {192
\pi^2}} {k^{3 \over 2} \over {k_B^2 \sqrt{k_b k_0}}} \sqrt{  {\langle b_0^2
\rangle \over \langle B_{ff0}^2 \rangle } \, F_{\delta}(k, k_0, k_b)} \,.
\eqno(70) $$ 

Although $k_b \gg k_0$   within our assumption of scale  separation,  we 
may  take the limit $k_b \to k_0$ to make a connection with \S 3.2.1. In this
limit, we  recover eq. (37) with an overall factor of 2 difference due to the
definitions of $\langle B_{ff0}^2 \rangle$, $\langle b_0^2 \rangle$, and 
$\langle B_{0}^2 \rangle$. 

Given the linear relation between the generated compressible modes and the
magnetic field perturbations about the force-free background, (i.e., $\delta
\propto b$), it is not surprising that the range of excited modes of $\delta$
is very narrow ($k_b - k_0 \le k \le k_b+k_0$) for the case  of a delta
function spectrum for $b$. In this scenario only structure with $k \simeq
k_b$ may form.   

Although this scenario is not very realistic, we briefly discuss the
formation of non-linear structure for the following limiting cases: $k_b =
k_B$ and $k_b \ll k_B$.
 For $  k_b = k_B$, we set  $\Delta (k_B,t_{nl}) = c(k_B) \tau(k_B,t_{nl})
\sqrt{ 32 \pi^4 k^3 /V} = 1$ and find that $t_{nl}$ satisfies
  $$\left({t_{nl} \over t_{\rm rec}}\right)^{1 \over 6} (1-\gamma_{\delta}) =
1 + {1 \over 6}{\rm ln}\left({t_{nl} \over t_{\rm rec}}\right) \ ,
\eqno(71)$$  where $\gamma_{\delta} = 3.8 \times 10^{-2} \sqrt{k_0 / k_B}
\sqrt{\langle  B_{ff0}^2\rangle\ /\langle b_0^2 \rangle}$.  Since  $t_{nl} \to
t_{\rm rec}$ as $\gamma_{\delta} \to 0$, the smaller $k_0 /k_B$ the earlier
non-linear structure on scale $k_B$ can form. 

As $k_b/k_B$ decreases from unity,  the cutoff density spectrum becomes $k_b
+k_0$ instead of $k_B$. For  $k_b \ll k_B $,
 we can approximate $m\simeq 5$ in eq. (60a) and write: $$ \left({t_{nl}(k_b)
\over t_{\rm rec}}\right)^{2 \over 3} \simeq 1.5  \left({k_0 \over
k_b}\right)^{1 \over 2} \left({k_B\over k_b}\right)^2  \left( {\langle 
B_{ff0}^2\rangle\ \over \langle b_0^2 \rangle} \right)^{1 \over 2} \ .
\eqno(72) $$ The time dependence in this case is similar  to the examples
discussed in \S 3.2.1.

The narrow range of excited modes  limits the applicability of these results
to physically interesting scenarios. We therefore move on to the more
physically relevant case of power law spectra for ${\tilde b}(k)$.

\smallskip  \centerline{\it 4.2.2 Power Law Spectra for ${\tilde b}^2(k)$}
\smallskip

The more general case of a power law spectrum of magnetic field
perturbations can be studied if we assume
 that, for $k_{\rm min} \le k \le k_{\rm max}$,  $$ {\tilde b}^2(k)\equiv  C 
\; k^{q} \,. \eqno(73) $$  where $C =  (q + 3) \; \langle b_0^2 \rangle \;
(k_{\rm max}^{q+3} - k_{\rm min}^{q+3})^{-1}$. For $k_{\rm min} \ll k_{\rm
max}$, $C \simeq (q + 3) \; \langle b_0^2 \rangle \; k_{\rm max}^{-q-3}$.  In
Appendix D, we show   that for this Ansatz (eq.\ (D8))
 \eq  k^2 {\dot {\tilde v}}_{{\p}\rm rms}^2 ({k},t_{\rm rec})  \simeq {
\alpha^2 V \over 4 (4\pi)^4} \; \langle B_{ff0}^2 \rangle \; \langle b_0^2
\rangle \;
 { k_{\rm max}^5 \over \, k_0^3 k  } \, (q+3) \, F_q(k,k_0,k_{\rm max})
\eqno(74)$$ 
where $F_q(k,k_0,k_{\rm max})$ is given by eq. (D9).
Therefore,  we find that  
$$  P_{q}(k,t) \simeq \left({25 \over 192}\right)^2
\; {r (q+3) \over 2 \pi} \; { k_{\rm max}^5 \over k_B^4 k_0^3 k} \;
F_q(k,k_0,k_{\rm max}) \; \tau(k,t)^2 , \eqno(75) $$   and $$ \Delta_{q}
(k,t) \simeq {25  \over 96} \sqrt{{r (q+3) F_q(k,k_0,k_{\rm max}) \over 2}}
\left({k_{\rm max} \over k_0}\right)^{3 \over 2}\;{k_{\rm max} k  \over
k_B^2} \, \tau (k,t) \ \ , \eqno(76) $$   where  $$ r \equiv { \langle
b_0^2  \rangle \over \langle B_{ff0}^2  \rangle} \ . $$ 

In Fig. 2, we plot $F_q(k,k_0,k_{\rm max})$ for different choices of $q$
with fixed $k_0/k_{\rm max}= 10^{-4}$. If we take the limit  $k_0/k_{\rm
max} \to 0 $ in eq. (D5) (note that the limits of integration depend on
$k_0$), $F_q$ can be approximated by 
$$F_q \simeq 16 \left({k_0 \over k_{\rm
max}}\right)^3 \; \left( {k \over k_{\rm max}} \right)^{q+5} \ , $$
as it is clearly shown in Fig. 2. 
We can, therefore, write 
$$  P_{q}(k,t) \simeq \left({25 \over 48}\right)^2 \; {r (q+3) \over 2 \pi}\;
 { k_{\rm max} \over k_B^4 } \; \left({k \over k_{\rm
max}}\right)^{q+4}  \tau(k,t)^2 , \eqno(77) $$   and $$ \Delta_{q} (k,t)
\simeq {25  \over 24} \; \sqrt{{r (q+3) \over 2}} \; \left({k_{\rm max} \over
k_B}\right)^2 \; \left({k  \over k_{\rm max}}\right)^{q + 7 \over 2} \, \tau
(k,t) \ \ . \eqno(78) $$  

The spectrum of generated density perturbations in this case depends both on
the magnetic field spectral index ($P_{q} \propto k^{q+4}$, for $k \ll
k_B$) and the ultraviolet cutoff $k_{\rm max}$.  The amplitude of
density perturbations  is determined by $k_B$, $k_{\rm max}$, and $r$, the
ratio between the energy density in magnetic perturbations relative to
that of the background field. We have concentrated on $q \ge -3$ such that
the dependence on $k_{\rm min}$ is negligible. The results can be easily
extended for cases in which $k_{\rm min}$ plays an important role. 

As an example, we choose the case of $q= 0$ and plot $\Delta_{0} (k,t)/ 
\sqrt{r}$ at different redshifts in Fig.\ 3, asssuming $  k_B= k_{\rm max} $,
$k_0 = 10^{-4} k_{\rm max}$, and $ k_{\rm min} = 10^{-3}  k_{\rm max}$. For 
$k \ll  k_{\rm max}$,  $P_{q}
\propto k^4$ and the peak power is at $k \sim  k_{\rm max}= k_B$.  In this
example, the behavior at small $k$ is too steep to fit large scale structure
observations and structure will be formed primarily around the scale $k_{\rm
max}$ or $ k_B$. To broaden the influence of magnetic fields on the range of
linear structure a spectral index $q \sim -3$ is necessary (see Ratra
(1992a,b), for an example of $q=-3$ primordial magnetic field spectrum).

We now discuss some examples within the large  parameter space ($k_{\rm
max}$, $k_0$, $k_B$, and $r$) that are relevant to the formation of
structure in the universe. As long as $q \ge -3$,
 the first scales to become
non-linear have $k \sim {\rm min}( k_{\rm max},k_B)$. We again define
$t_{nl}(k)$, such that $\Delta_{q} (k,t_{nl}) =1$. 
For  $ k_{\rm max} +k_0 >
k_B$, $k_B$ is the largest mode to be excited. Setting $\Delta_{q}
(k_B,t_{nl}) =1$ also leads to eq.\ (71) but with 
$$\gamma_{q}  \simeq {3.8
\times 10^{-2} \over \sqrt{r (q+3)}} \left(k_{\rm max}\over k_B
\right)^{q +3 \over 2} \ .$$ 
Since as $\gamma_{q} \to 0$, $t_{nl} \to t_{\rm rec}$, for $ k_{\rm max} +k_0 >
k_B$ the closer $k_{\rm
max}$ is to $k_B$, the earlier structures on scales $k_B$ can form. 

To simplify our discussion, we set $q=0$ and $r=1$ and vary $k_{\rm max}$,
$k_B$, and $k_0$ (different choices of $q$ and $r$ can be accommodated by
making the appropriate rescaling of $k_{\rm max}$). We find that for
$k_{\rm max} \simeq k_B$, the first non-linear objects could form as early as
redshift $z_{nl}(k_B) \simeq 300$ and the scale of non-linearity  today would
be  $k_{nl}(t_0) \simeq 0.15 \; k_B$. If we set $k_{nl}(t_0) = 2 \pi /
8$ Mpc, then $k_{\rm max} \simeq k_B \simeq 2 \pi / 1 $ Mpc, which satisfies
eq.\ (10). Since the background field is a delta function the appropriate
bound on the parameters due to the upper limit on magnetic fields on Mpc scales
is given by  eq.\ (43) with $k_0$ substituted for $k_*$ (as long as $r \ll 1$).
In this example, eq. (43) is satisfied if $k_0 = 2 \pi / 2$ Mpc.  If instead,
we choose $k_{nl}(t_0) = 2 \pi / 2$ Mpc, then $k_{\rm max} \simeq k_B \simeq 2
\pi / 0.3 $ Mpc which satisfies eqs.\ (10) and (43) with $\langle B_0^2
\rangle^{1/2} \simeq 10^{-8}$ G and $\bar B(l_g,t_0) \simeq 10^{-10}$
G (for $k_0 \simeq 2 \pi / 2$ Mpc).

The time at which modes with $ k=k_B $ become non-linear, $t_{nl}(k_B)$,
increases as $k_{\rm max}/k_B$ grows from unity. For magnetic fields to play
a role in structure formation, $t_{nl}(k_B)$ must be less than the age of the
universe, $t_0$, which implies that $k_{\rm max} \lsim 6 k_B$.

As $k_{\rm max}/k_B$ decreases, so does $t_{nl}(k_B)$ up to $k_{\rm max}+ k_0
\simeq k_B$,  where the cutoff changes from $k_B$ to $k_{\rm max}+ k_0$. For $
k_{\rm max} + k_0 < k_B $,  we follow
$t_{nl}(k_{\rm max})$ using  our solution (eq. \ (60a)) in the limit
 $k \ll k_B$ ($m \simeq 5$) which implies:  
$$ \left({t_{nl}(k) \over t_{\rm
rec}}\right)^{2 \over 3} \simeq {1.5 \over \sqrt{r (q+3)}} \left({k_B
\over k_{\rm max}}\right)^{ 2 } \left({k_{\rm max}
\over k}\right)^{q +7 \over 2} \,.
\eqno(79) $$ 
From eq.\ (79), requiring $t_{nl}( k_{\rm max}) \lsim
t_0$ leads to the constraint $ k_{\rm max}  \gsim 10^{-2} k_B$. This estimate
helps define the range for which magnetic fields can make non-linear
structures, i.e., $10^{-2} k_B \lsim k_{\rm max} \lsim 6 k_B$. For example, we
can choose $ k_{\rm max} \simeq 0.1 k_B$, which gives $z_{nl}(k_{\rm max})
\simeq 7$. Choosing $k_{\rm max} \simeq 2 \pi / 0.8$ Mpc,  then $k_B \simeq 2
\pi /80 $ kpc and all the observational constraints are satisfied, with
$\langle B_0^2 \rangle^{1/2} \simeq 3 \times 10^{-9}$ G and $\bar
B(l_g,t_0) \simeq 10^{-10}$ G.  On the other hand, if $k_{\rm max} \simeq
10^{-2} k_B \simeq 2 \pi / 8$ Mpc,  then $t_{nl}( k_{\rm max}) \simeq t_0$ and
$k_B = 2 \pi / 0.08$ Mpc which satisfies eq.\ (10) but violates eq.\ (49). 

We see that with a more realistic choice of power spectra for magnetic
fields at recombination, non-linear structures in a cosmologically relevant
range of scales can be formed at reasonable redshifts for field strengths
below the observational upper limits.  Although the number of parameters in
this discussion is large, the  physically relevant range is quite limited. As
we improve our understanding of the evolution of magnetic fields  from the
early universe up to  the epoch of galaxy formation, the parameter space will
become  more constrained. For instance, in the absence of primordial
vorticity, magnetic field perturbations are damped up to the Silk damping
scale and $k_{\rm max} \lsim 2 \pi / 10$ Mpc (Jedamzik, Katalinic, \& Olinto
1996).

\bigskip \centerline{\bf 5. Conclusions} \medskip

We have studied the effects of magnetic fields present at recombination on
the origin and evolution of density perturbations and  peculiar velocities. 
We find there are generic features of the generated density
perturbations which are largely independent of the assumed spectrum of the
primordial magnetic field. The first conclusion we can draw is that, within
our linear perturbation approach,   magnetic fields cannot explain
the observed galaxy power spectrum on large scales, since the generated
spectrum is too steep for small $k$. To reproduce the observed $P \propto k$
spectrum on large scales, we need ${\tilde b}^2(k) \propto k^{-3}$ for small
$k$ with amplitudes that violate the observational constraints on $\langle
B_0^2 \rangle$.

Another generic feature is the cutoff introduced by the magnetic Jeans
length. This cutoff limits the amplitude of the power spectrum for any choice
of magnetic field strength. As the magnetic field strength increases, the
amplitude for a given density perturbation mode rises, but, simultaneously,
the magnetic Jeans cutoff moves to smaller $k$. The net effect is that the
peak amplitude for the resulting density power spectrum does 
not necessarily increase
as one increases $\langle B_0^2 \rangle$. 

Since the generated spectrum falls sharply at small $k$ and is cut off at
large $k$, magnetic fields generally produce a peak in the density spectrum
over a narrow range of wavenumbers. For this peak to be of relevance to the
formation of structure, the amplitude $\Delta(k,t_0)$ must be $\gsim 1$ for
scales $2 \pi k^{-1} \lsim 8 $ Mpc. The peak amplitude is sensitive to the
assumed spectrum for the primordial magnetic field; the smaller $k_{\rm max}$,
the stronger the variance in the density perturbations. In particular, if
$k_{\rm max} \lsim k_B$, density variances well above unity can be obtained
with relatively small magnetic fields. Depending on $k_B$ and $k_{\rm max}$,
objects from galaxy scales down to a first generation of massive stars can be
formed. As the variance reaches unity our calculations break down, and
another conclusion can be drawn: when magnetic fields are important, their
effects are mostly non-linear in nature. Our linear calculations can,
however, be used to estimate the epoch of non-linear collapse of different
mass scales.

In the present work, we have focused on a purely baryonic universe. In a
subsequent paper, we study the evolution of density perturbations when
non-baryonic dark matter is the dominant component of the universe. When we
include non-relativistic (cold) dark matter, we find that the amplitude of
the density perturbations decreases as $\Omega_{\rm baryon}$ decreases for a
fixed magnetic field strength. On the other hand, the perturbations become
unstable for all wavenumbers: since cold dark matter does not couple to the
magnetic fields, no damped oscillatory modes can be sustained. 
In this case, the perturbation
amplitude $\Delta(k)$ flattens out for large $k$ rather than being cutoff at
$k_B$. Baryons will still show some resistance to clumping on small scales
due to the magnetic field; this may segregate baryons from dark matter,
introducing a source of bias on small scales.

In the case of hot dark matter the growth of baryon perturbations on small
scales is slower due to neutrino free-streaming. Therefore, the initial
perturbations need to be much larger for the final variance to be greater
than 1. Clearly, the hot dark matter scenario would greatly benefit from a
peak in the variance at large $k$ so that structures can form on scales
smaller than the neutrino free-streaming length and populate the problematic
empty voids. 

Finally, we conclude by noting that magnetic fields most likely play a
dynamical role in the formation of galaxies and clusters of galaxies even if
the original perturbations were caused by other sources.

\bigskip\bigskip \centerline{\bf Acknowledgements:} We thank J.\ Frieman, A.\
Stebbins, S.I.\ Vainshtein, and I.\ Wasserman for useful discussions. This
work was supported in part by a NASA Astrophysics Theory grant to the Univ.\
of Chicago and NASA grant NAG 5-2788 at Fermilab.

\vfill\eject

\centerline{\bf Figure Captions} \medskip

\noindent Fig.\ 1: The transfer function $\tau(k,t)$ plotted versus $k/k_B$
for redshifts $z= 0,1,5,10,100,$ and 1000. 
We plot $\tau (k,t/t_{\rm rec})$ for $k > k_{\rm rec}$ and $\tau
(k, t/t_{\rm enter})$ for $k<k_{\rm rec}$ 
with $t_{\rm enter}(k) = t_{\rm rec} (k_{\rm rec} / k)^3$.
(We set $k_{\rm rec} = 0.01 k_B$.) 

\smallskip \noindent Fig.\ 2: The function $F_q$ for the power law Ansatz as a
function of  $k/k_{\rm max}$, for $k_0 = 10^{-4} k_{\rm max}$, $k_{\rm min} = 
10^{-3} k_{\rm max}$, and $q=-3, -2, -1, 0, 1, 2$.

\smallskip \noindent Fig.\ 3: The variance for the power law Ansatz with $q=0$, 
$\Delta_{0}/ \sqrt{r}$,  as a function of $k/k_B$ for
redshifts $z= 0,1,5,10,100,$ and 1000. (We set $k_{\rm max} = 10^3 k_{\rm min}=
10^4 k_0$ as in Fig.\ 2.)

\vfill\eject \centerline{\bf REFERENCES} \bigskip \ref Barrow, J.\ \ 1976,
MNRAS, 175, 359 \ref Cattaneo, F.\ \ 1994, ApJ, 434, 200 \ref Cattaneo, F., \&
Vainshtein, S.I.\ \ 1991, ApJ, 376, L21  \ref Chakrabarti, S.K., Rosner, R.,
\& Vainshtein, S. 1994, Nature, 368, 434 \ref Cheng, B., \& Olinto, A. V.\ \
1994, Phys.\ Rev.\ D, 50, 2421  \ref Cheng, B., Schramm, D., \& Truran, J.
1994, Phys. Rev. D, 49, 5006
 \ref Collins, C. A., Nichol,
R.C., \& Lumsden, S. L. \ \ 1992, MNRAS, 254, 295 
\ref Dolgov, A.D., 1993, Phys. Rev. D, 48,2499
\ref Dolgov, A.D., \& Silk, J. 1993, Phys. Rev. D, 47, 3144
\ref Efstathiou, G.,  et
al.\ \ 1990, MNRAS, 247, 10P \ref Fisher, K. B.,  et al.\ \ 1993, ApJ, 402,
42 \ref Ganga, K., Cheng, E., Meyer, S., \& Page, L.\ \ 1993, ApJ,
410, L57 \ref Geller, M., \& Huchra, J.\ \ 1989, Science, 246, 897 
\ref Grasso, D., \& Rubinstein, H.R. \ 1995, Astropart. Phys. 3, 95 
\ref Gruzinov, A.V., \& Diamond,
P.H.\ \ 1994, PRL, 72, 1651 \ref
Hogan, C.J.\ \ 1983, PRL, 51, 1488 
\ref Jedamzik, K., Katalinic, V., \& Olinto, A. \ 1995, private communication
\ref Kernan, P., Starkman, G., \& Vachaspati, T. \ 1995, CWRU-P10-95
\ref Ko, C.M., \& Parker, E.N.\ \ 1989, ApJ,
341, 828 \ref
Kraichnan, R.H., \& Nagarajan, S.\ \ 1967, Phys.\ Fluids, 10, 859 \ref
Kronberg, P. P.\& Simard-Normandin,  \ 1976, Nature, 263, 653     \ref Kronberg,
P. P., \& Perry, J. J. \ \ 1982,  ApJ, 263, 518 \ref Kronberg, P. P., Perry,
J. J., \& Zukowski, E. L. H. \ \ 1992,  ApJ, 387, 528 \ref Kronberg, P. P.\ \
1994,  Rep.\ Prog.\ Phys., 57, 325 
\ref Kulsrud, R.M\ \ 1988, in Proc. Varenna Conf. Plasma Astrophys., ed. R. 
Gutenne (ESA SP-285)
 \ref Kulsrud, R.M., \& Anderson, S.W.\ \ 1992, ApJ, 396, 606
\ref Kulsrud, R.M. \ 1995, private communication
\ref Lee, S., Olinto, A., \& Sigl, G. \ \ 1995, ApJ, 455, L21
\ref Maddox, S. J., et al.\ \ 1990, MNRAS,
246, 433 \ref Moffatt, H.K.\ \ 1978,  Magnetic Field Generation in
Electrically-Conducting Fluids  (Cambridge: Cambridge Univ.\ Press)
 \ref
Parker, E.N.\ \ 1979, {\it Cosmical Magnetic Fields} (Oxford: Oxford Univ.\
Press) \ref Parker, E. N.\ \ 1992, ApJ, 401, 137 \ref
Peebles, P. J. E.\ \ 1980,   The Large--Scale Structure of The  Universe
(Princeton: Princeton Univ.\ Press) \ref Peebles, P. J. E.\ \ 1993,  
Principles of Physical Cosmology (Princeton: Princeton University Press) 
\ref
Quashnock, J., Loeb, A., \& Spergel, D.N.\ \ 1989, ApJ, 344, L49 \ref Ratra,
B.\ \ 1992a, ApJ, 391, L1
\ref Ratra,
B.\ \ 1992b, Phys. Rev. D, 45, 1913  
\ref Rosner, R., \& DeLuca,
E.E.\ \ 1989, in {\it The Center of the Galaxy}, ed.\ M.\ Morris
(Dordrecht:
Kluwer), p.\ 319
\ref Smoot, G.,  et al.\ \ 1992, ApJ, 396,
L1  \ref Taylor, J.B.\ \ 1974, PRL, 33, 1139 \ref 
Turner, M.S., \& Widrow, L.M.\ \ 1988, Phys.\ Rev.\ D, 30, 2743 
\ref Vainshtein,
S.I., \& Rosner, R.\ \ 1991, ApJ, 376, 199
\ref Vaschaspati, T.\ \ 1991, Phys.\ Lett.\ B, 265, 258  \ref Wasserman, I.\ \
1978, ApJ, 224, 337 \ref Wolfe, A.M. \ \ 1988 in  QSO Absorption Lines:
Probing the Universe, ed.\ Blades, C.J., et al.\ (Cambridge: Cambridge Univ.\
Press), 297 \ref Wolfe, A.M., Lanzetta, K.M., \& Oren, A.L.\ \ 1991, ApJ, 388,
17
\ref Woltjer, L.\ \ 1958, Proc.\
Nat.\ Acad.\ Sci., 44, 489 \ref Zweibel, E.G.\ \ 1988, ApJ, 329, 384.

\vfill\eject

\bigskip \centerline{\bf Appendix A}  \medskip 

In this Appendix, we derive eq.\ (34), apply them to the  two
Ansatze for ${\tilde {\B}}({\k})$, and derive eqs. (36) and (45). We begin
with eq.\ (31),

\eq {\x \cdot {\dot{\v}}({\bf x},t_{\rm rec})}=\alpha {\x \cdot[(\x\times
{\B}_0) \times {\B}_0]} \,, \eqno(A1) \en where $\alpha \equiv 1/ 4 \pi
\rho_0 R^2_{\rm rec}$. By using the Fourier expressions for ${\v}$ and
${\B}_0$, we obtain

\eq i{\k}\cdot {\dot {\tilde {\v}}}({\k},t_{\rm rec}) = \alpha\int d^3 {\k}_1
\ {\k} \cdot\{ {\k}_1 [{\B}({\k}_1)\cdot {\B} ({\k}-{\k}_1)]
-{\B}({\k}_1)[{\k}_1\cdot {\B}({\k}-{\k}_1)]\} \,, \eqno(A2) \en where ${\B}
\equiv {\tilde {\B}}$ throughout this Appendix. If we choose our coordinate
system in such a way that ${\k}$ is along the $z$--axis, 

\eq \int d^3 {\k}_1 = 4\pi \int dk_1 k_1^2 \int d{\mu} \ , \en where ${\mu}
\equiv \cos {\theta}$ with ${\theta}$ the angle between ${\k}_1$ and
the $z$--axis. The integration range in eq.\ (A2) should be taken carefully 
since
the integrand had  $\delta ({\k}-{\k}_1-{\k}_2) $ before integrated over
${\k}_2$ with the condition of $k_{\rm min} \leq k_i \leq k_{\rm max}$ for
$i=1,2$. In other words, there is a constraint on the angle to be integrated
over, depending on the magnitude of $k_1$. By taking this constraint 
into account, the integration range of eq.\ (A2) depends on the magnitude of
$k$, and can be shown to be

\medskip\noindent (1) For $0\leq k\leq k_{\rm min}$:

\eq\eqalignno{ \int dk_1 d{\mu} &= \int _{k_{\rm min}}^{k+k_{\rm min}} dk_1
\int_{-1}^{k^2+k^2_1-k^2_{\rm min} \over 2kk_1}d\mu + \int _{k+k_{\rm
min}}^{k_{\rm max}-k} dk_1 \int_{-1}^1 d\mu &\cr & &\cr &+ \int _{k_{\rm
max}-k}^{k_{\rm max}} dk_1  \int^1_{k^2+k^2_1-k^2_{\rm max} \over 2kk_1}d\mu
\ . &(A3)\cr} \en (2) For $k_{\rm min} < k < k_{\rm max}$:

\eq\eqalignno{ \int dk_1 d{\mu} &= \int _{k_{\rm min}}^{k-k_{\rm min}} dk_1
\int_{-1}^1 d\mu + \int _{k-k_{\rm min}}^{k+k_{\rm min}} dk_1
\int_{-1}^{k^2+k^2_1-k^2_{\rm min} \over 2kk_1}d\mu &\cr & &\cr &+ \int
_{k+k_{\rm min}}^{k_{\rm max}-k} dk_1 \int_{-1}^1 d\mu + \int _{k_{\rm
max}-k}^{k_{\rm max}} dk_1 \int^1_{k^2+k^2_1-k^2_{\rm max} \over 2kk_1}d\mu \
. &(A4)\cr} \en (3) For $k_{\rm max} \leq k \leq k_{\rm max}+k_{\rm min}$:

\eq \int dk_1 d{\mu} =\int _{k_{\rm min}}^{k-k_{\rm min}} dk_1
\int^1_{k^2+k^2_1-k^2_{\rm max} \over 2kk_1}d\mu +\int _{k-k_{\rm
min}}^{k_{\rm max}} dk_1 \int_{k^2+k^2_1-k^2_{\rm max}\over
2kk_1}^{k^2+k^2_1-k^2_{\rm min} \over 2kk_1}d\mu \ . \eqno(A5)\en (4) For
$k_{\rm max}+k_{\rm min} \leq k \leq 2 k_{\rm max}$:

\eq \int dk_1 d{\mu} =\int ^{k_{\rm max}}_{k_{\rm min}} dk_1
\int^1_{k^2+k^2_1-k^2_{\rm max} \over 2kk_1}d\mu\ . \eqno(A6) \en If we take
the ensemble average of both sides of eq.\ (A2), and use eq.\ (29), we
obtain   $$ \langle {\k}\cdot {\dot {\tilde {\v}}}({\k},t_{\rm rec})\rangle=0
\,. $$  To calculate ${\dot {\tilde v}}_{\rm rms}(k,t_{\rm rec})$, we take
the square of both sides of eq.\ (A2) and obtain  \eq\eqalignno{  |{\k}\cdot
{\dot {\tilde {\v}}}({\k},t_{\rm rec})|^2  =\alpha^2 &\int d^3 {\k}_1 d^3
{\k}_2 \{(-{\k}\cdot{\B}^*({\k}_1))[{\k}_1\cdot{\B}^*({\k}-{\k}_1)]
+({\k}\cdot {\k}_1) [{\B}^*({\k}_1)\cdot {\B}^* ({\k}-{\k}_1)]\} & \cr
&\times \{(-{\k}\cdot{\B}({\k}_2))[{\k}_2\cdot{\B}({\k}-{\k}_2)] +({\k}\cdot
{\k}_2) [{\B}({\k}_2)\cdot {\B} ({\k}-{\k}_2)]\}, &(A7)\cr} \en We then take
the ensemble average of both sides of eq.\ (A7), and express the fourth-order
correlation of $B$ as a sum of a product of second order correlations by
assuming a Gaussian statistics for $B$. Since each fourth-order correlation
gives three products of second-order ones, we obtain a total of twelve terms
on the right hand side of eq.\ (A7). Applying eq.\ (29) to these twelve terms
yields terms proportional to $\delta ({\k})$, $\delta({\k}_1-{\k}_2)$, and
$\delta ({\k}-{\k}_1-{\k}_2)$, respectively. Four of these terms (which
contain $ \delta ({\k})$) vanish upon integration and they will not be
written here; retaining the remaining eight terms leads to the expression

\eq\eqalignno{
 \langle |{\k}\cdot {\dot {\tilde {\v}}}({\k},t_{\rm rec})|^2 \rangle 
=\alpha^2 &\int d^3 {\k}_1 d^3 {\k}_2 \bigl\{ \langle ({\k}\cdot
{\B}^*({\k}_1) {\k}\cdot {\B}({\k}_2) \rangle 
 \langle {\k}_1\cdot{\B}^*({\k}-{\k}_1) {\k}_2 \cdot {\B}({\k}-{\k}_2)
\rangle  &\cr &+({\k}\cdot{\k}_1)( {\k}\cdot{\k}_2 ) \langle
B_i^*({\k}_1)B_j({\k}_2) \rangle 
 \langle B_i^*({\k}-{\k}_1)B_j({\k}-{\k}_2) \rangle  \cr &-({\k}\cdot {\k}_2)
\langle {\k}\cdot{\B}^*({\k}_1) B_i({\k}_2) \rangle 
 \langle {\k}_1\cdot{\B}^*({\k}-{\k}_1) B_i({\k}-{\k}_2) \rangle  &\cr
&-({\k}\cdot {\k}_1) \langle {\k}\cdot{\B}({\k}_2) B_i^*({\k}_1) \rangle 
 \langle {\k}_2\cdot{\B}({\k}-{\k}_2) B_i^*({\k}-{\k}_1) \rangle  &\cr &+
\langle ({\k}\cdot {\B}^*({\k}_1) {\k}_2\cdot {\B}({\k}-{\k}_2) \rangle 
 \langle {\k}\cdot{\B}({\k}_2) {\k}_1 \cdot {\B}^*({\k}-{\k}_1) \rangle  &\cr
&+({\k}\cdot{\k}_1)( {\k}\cdot{\k}_2) \langle B_i^*({\k}_1)B_j({\k}-{\k}_2)
\rangle 
 \langle B_i^*({\k}-{\k}_1)B_j({\k}_2) \rangle  &\cr &-({\k}\cdot {\k}_2)
\langle {\k}\cdot{\B}^*({\k}_1) B_i({\k}-{\k}_2) \rangle 
 \langle {\k}_1\cdot{\B}^*({\k}-{\k}_1) B_i({\k}_2) \rangle  &\cr
&-({\k}\cdot {\k}_1) \langle {\k}\cdot{\B}({\k}_2) B_i^*({\k}-{\k}_1) \rangle 
 \langle {\k}_2\cdot{\B}({\k}-{\k}_2) B_i^*({\k}_1) \rangle \bigr\} &\cr
&\equiv k^2 {\dot {\tilde v}}_{{\p}\rm rms}^2 ({\k},t_{\rm rec}) \ , &(A8)
\cr} \en  where ${\k}\cdot {\dot {\tilde {\v}}}({\k},t_{\rm rec}) \equiv
k{\dot{\tilde {v}}}_{\p}({\k},t_{\rm rec})$. The integral of these eight
terms is given by (using eq.\ (29)),

\eq\eqalignno{ k^2 {\dot {\tilde v}}_{{\p}\rm rms}^2 ({k},t_{\rm rec})
={\alpha^2 V k^2  \over (4\pi)^2} &\int_{k_{\rm min}}^{k_{\rm max}}dk_1 k_1^2
\int^1_{-1}d{\mu}  {B^2(k_1)B^2(|{\k}-{\k}_1|) \over |{\k}-{\k}_1|^2} &\cr &
&\cr &\times \bigl \{2 k^3 k_1 \mu + k^2 k_1^2 ( 1- 5\mu^2) + 2 k k_1^3 \mu^3
\bigr \} \,, &(A9)\cr} \en where the identity $\delta({\k}=0) =V/(2\pi)^3$
has been used. Eqs. (A8) and (A9) give eq. (34).

For the delta function Ansatz, $B^2(k) = \langle B_0^2 \rangle
\delta(k-k_{*})/ 4 \pi k^2$, and therefore $$  k^2 {\dot {\tilde v}}_{{\p}\rm
rms}^2 ({k},t_{\rm rec}) = { \alpha^2 V \langle B_0^2 \rangle^2 \over (4
\pi)^4}  {k^3 \over k_{*}^2} \,. \eqno(A10) $$
 Eq.\ (36) follows from this result.

We now evaluate ${\dot {\tilde v}}_{{\p}\rm rms}^2 $ for the case of a power
law spectrum for $\tilde{B}^2(k)$, as in eq.\ (44). For simplicity, we assume
$k_{\rm min} = 0$. The resulting integral can be solved analytically if we
use $k/k_{\rm max}$ as a small parameter. In such an expansion, the leading
order term turns out to be $(k/k_{\rm max})^3$ for $q=-2$, and $(k/k_{\rm
max})^4$ for $q=-1,$ 0, 1, 2, 3, 4, and 6. Finally, we find the following
solution for the $q \ge -1$ case:

\eq k^2 {\dot {\tilde v}}_{{\p}\rm rms}^2 ({k},t_{\rm rec})= {\alpha^2 V\over
(4\pi)^4}A^2 k_{\rm max}^{2q+3} k^4 {22\over 15(2q+3)} \biggl[ 1+ O \biggl({
k\over k_{\rm max}} \biggr)\biggr]. \eqno(A11) \en  Therefore,

\eq
 \langle |{\k}\cdot {\dot {\tilde {\v}}}({\k},t_{\rm rec})|^2 \rangle  = k^2
{\dot {\tilde v}}_{{\p}\rm rms}^2 ({k},t_{\rm rec}) \simeq  {\alpha ^2 V\over
(4\pi)^4} \langle B_0^2 \rangle^2 { k^4\over k_{\rm max}^{3}} {22 (q+3) \over
15(2q+3)}, \eqno(A12) \en 
where we used $ \langle B_0^2 \rangle = 4 \pi \int ^{k_{\rm max}}_{k_{\rm min}}
dk k^2
\; Ak^q \simeq Ak_{\rm max}^{q+3}/(q+3)$; this leads directly to eq.
(45).

 \bigskip \centerline{\bf Appendix B}  \medskip

 In this Appendix, we demonstrate that $\langle B_{\rm rec} {\tilde
v}_{\p}(t) \rangle = \langle B_{\rm rec} {\tilde v}_{\t}(t) \rangle = 0$. We
begin with the evolution equation for ${\tilde v}_{\p}$, obtained through
eq.\ (54)  $$
 k \left(t^2 \partial _{tt} +2 t \partial _t -{4\over9}\right) \ {\tilde
v}_{\p}({\k},t) =\beta \int d^3 {\bf k}_1 d^3 {\bf k}_2 \ F( {\k},{\k}_1,{\bf
k}_2,\tilde {\B}({\k}_1),\tilde{\B}({\k}_2), {\tilde {\v}}({\bf k}-{\bf
k}_1-{\bf k}_2,t)) 
 \eqno(B1) $$ With the use of the Green function for this equation, \eq
G(t,t')={3\theta (t-t')\over 5kt'}\biggl[\biggl({t'\over t}\biggr)^{-{1\over
3}} -\biggl({t'\over t}\biggr)^{4\over 3}\biggr] \,, \eqno(B2) \en  (where
$\theta$ is the step function), one can solve eq.\ (B1) formally, given the
initial conditions $$ {\tilde v}_{\p}(k,t_{\rm rec})=0 $$ and  $$
{\partial}_t {\tilde v}_{\p}(k,t)|_{t=t_{\rm rec}} = {\dot {\tilde
v}}_{\p}(k,t_{\rm rec}) \,. $$ This Green's function solution is given by
\eq\eqalignno{ {\tilde v}_{\p}(k,t) &={3 t_{\rm rec}{\dot {\tilde
v}}_{\p}(k,t_{\rm rec})\over 5} \biggl[\biggl({t_{\rm rec}\over t}
\biggr)^{-{1 \over 3}} -\biggl({t_{\rm rec}\over t}\biggr)^{4 \over 3}\biggr]
+ &\cr & {\beta}\int ^t_{t_{\rm rec}} dt' \int d^3 {\bf k}_1 d^3 {\bf k}_2 
G(t,t') F( \k,\k_1,\k_2, \tilde {\B} ({\k}_1),\tilde{\B}({\k}_2), {\tilde
{\v}}({\k}-{\k}_1-{{\k}_2},t')) \,. &(B3)\cr} \en This equation can be solved
by iteration, yielding a Born-series solution, for which the zeroth-order
iterate ${\tilde {v}}_{\p}^0(k,t)$ is given by \eq {\tilde {v}}_{\p}^0(k,t)
={3 t_{\rm rec}{\dot{\tilde {v}}}_{\p}(k,t_{\rm rec})\over 5}
\biggl[\biggl({t_{\rm rec}\over t} \biggr)^{-{1 \over 3}} -\biggl({t_{\rm
rec}\over t}\biggr)^{4 \over 3}\biggr] \,. \eqno(B4) \en

We first note that $ \langle {\tilde v}_{\p}^0(k,t)B_{\rm rec} \rangle =0$.
This result is obtained simply by noting that ${\dot{\tilde
v}}_{\p}^0(k,t_{\rm rec})$ is quadratic in $B_{\rm rec}$ (cf.\ eq.\ (55)), so
that ${\tilde v}_{\p}^0(k,t) B_{\rm rec}$ is an odd product of $B_{\rm rec}$.
Hence, as long as $B_{\rm rec}$ is a random variable governed by Gaussian
statistics, $\langle {\tilde v}_{\p}^0(k,t) B_{\rm rec} \rangle$ must vanish
since all odd moments of a Gaussian random variable vanish.

It is now straightforward to demonstrate that $ \langle {\tilde
v}_{\p}(k,t)B_{\rm rec} \rangle =0$ to all orders by simply computing the
next-order iterates: At the $n^{\rm th}$ iteration stage, ${\tilde
v}_{\p}^n(k,t)$ is the sum of even products of $B_{\rm rec}$ (because $F$ is
a quadratic form in $B_{\rm rec}$); hence, $ {\tilde v}_{\p}^n(k,t)B_{\rm
rec}$ must be always an odd product of $B_{\rm rec}$, and therefore $ \langle
{\tilde v}_{\p}^n(k,t)B_{\rm rec} \rangle =0$, as before. Thus, under the
assumption that the Born series expansion converges, we obtain a portion of
our desired result, namely that  $$ \langle {\tilde v}_{\p}(k,t)B_{\rm rec}
\rangle =0 \,. $$ It is now readily shown that the same result obtains for
the incompressible flow, ${\tilde v}_{\t}$, by simply repeating the above
calculation, but now projecting out the incompressible component. Thus, we
also obtain $ \langle {\tilde v}_{\t}(k,t)B_{\rm rec} \rangle =0$, and hence
we have the ultimately desired result that $$ \langle {\tilde v}(k,t)B_{\rm
rec} \rangle =0 \,. $$

\bigskip  \centerline{\bf Appendix C}  \medskip  In this Appendix, we show how
to obtain ${\tilde {\delta}}_{\rm rms}(k,t)$ from ${\tilde v}_{\rm rms
}(k,t)$ in the force-free case of \S 4. We start with eq.\ (55) in the main
text, which reads

\eq (\partial _{TT} + \partial _T -{4\over 9}){\tilde v}_{\p}({\k},T) ={\beta
\over k}G_i( {\bf k:l}) {\tilde v}_i({\bf l}, T)\,. \eqno(C1) \en Upon using
$T\equiv \ln t$, this is equivalent to

\eq ( t^2 \partial _{tt} +2 t \partial _t -{4\over 9}){\tilde v}_{\p}({\k},t)
={\beta \over k}G_i( {\bf k:l}) {\tilde v}_i({\bf l}, t)\,. \eqno(C2) \en The
{\it rms} solution to (C2) was shown to be given by eq.\ (58) in the main
text, which we rewrite as  $$ {\tilde v}_{\p , \rm rms}(k,t) ={\dot {\tilde
v}}_{\p , \rm rms}(k, t_{\rm rec}) V(t) \ , \eqno(C3) $$ where $$ V(t) \equiv
{{3 t_{\rm rec}}\over m} \biggl({t \over t_{\rm rec}}\biggr)^{-{1\over
2}}\biggl[\biggl({t \over t_{\rm rec}}\biggr)^{{m}\over 6}-\biggl({t\over
t_{\rm rec}}\biggr)^{-{m \over 6} }\biggr] , $$ ${\tilde v}_{\rm rms}(k,t)
\equiv \sqrt {<|{\tilde v}_{\p}({\k}, t)|^2>}$, and $m \equiv
{\sqrt{25-12ak^2}}$.

First, we multiply both sides of eq.\ (C2) by ${\tilde v}^*_{\p}({\k},t')$,
with $t' \ne t$, and then take an average to get the following equation:

\eq ( t^2 \partial _{tt} +2 t \partial _t -{4\over 9}) \langle {\tilde
v}_{\p}({\k},t){\tilde v}^*_{\p}({\k},t')\rangle ={\beta \over k}\langle G_i(
{\bf k:l}) v_i({\bf l}, t) {\tilde v}^*_{\p}({\k},t')\rangle\,. \eqno(C4) \en
Now, after repeating the same calculation that was used to get eq.\ (57) from
eq.\ (56), it can be shown that

\eq \langle G_i ({\k}:{\bf l}) {\tilde v}_i({\bf l},t) {\tilde
v}^*_{\p}({\k},t') \rangle =-{2 \over 3} \langle B_0^2 \rangle k^3 \langle
{\tilde v}_{\p}({\k},t ){\tilde v}^*_{\p}({\k},t')\rangle \,. \eqno(C5) \en
Using eq.\ (C5), we readily obtain the solution to eq.\ (C4), namely

\eq \langle {\tilde v}_{\p}({\k},t) {\tilde v}^*_{\p}({\k},t')\rangle
=\langle {\dot {\tilde v}}_{\p}({\k}, t_{\rm rec}){\tilde v}^*_{\p}({\k},t')
\rangle V(t) \,. \eqno(C6) \en Next, let us take the complex conjugate of both
sides of eq.\ (C2), multiply by ${\dot {\tilde v}}_{\p} ({\k},t_{\rm rec})$,
and then take an average; that gives  \eq \langle {\tilde v}^*_{\p}({\k},t)
{\dot {\tilde v}}_{\p}({\k},t_{\rm rec})\rangle =\langle |{\dot {\tilde
v}}_{\p}({\k}, t_{\rm rec})|^2 \rangle V(t)\,. \eqno(C7) \en From eqs.\ (C3)
and (C7), \eq \langle {\tilde v}^*_{\p}({\k},t) {\dot {\tilde
v}}_{\p}({\k},t_{\rm rec}) \rangle =\sqrt{\langle |{\dot {\tilde
v}}_{\p}({\k}, t_{\rm rec})|^2 \rangle} \sqrt{\langle | {\tilde v}_{\p}({\k},
t)|^2 \rangle}\,, \en or \eq \langle {\tilde v}^*_{\p}({\k},t') {\dot {\tilde
v}}_{\p}({\k},t_{\rm rec}) \rangle =\sqrt{\langle |{\dot {\tilde
v}}_{\p}({\k}, t_{\rm rec})|^2 \rangle} \sqrt{\langle | {\tilde v}_{\p}({\k},
t')|^2 \rangle}\,. \eqno(C8) \en Upon using eq.\ (C8) in eq.\ (C6)

\eq \langle {\tilde v}_{\p}({\k},t) {\tilde v}^*_{\p}({\k},t')\rangle
=\sqrt{\langle |{\dot {\tilde v}}_{\p}({\k}, t_{\rm rec})|^2 \rangle}
\sqrt{\langle | {\tilde v}_{\p}({\k}, t')|^2 \rangle} V(t)\,. \eqno(C9) \en
Finally, with the help of eq.\ (C3), eq.\ (C9) becomes

\eq \langle {\tilde v}_{\p}({\k},t) {\tilde v}^*_{\p}({\k},t')\rangle
=\sqrt{\langle | {\tilde v}_{\p}({\k}, t)|^2 \rangle} \sqrt{\langle | {\tilde
v}_{\p}({\k}, t')|^2 \rangle}\,. \eqno(C10) \en On the other hand, eq.\ (12)
in the main text gives us the following equation:

\eq ik {{\tilde v}_{\p}({\k},t) \over R(t)}=-{\partial \over\partial
t}\,{\tilde \delta} ({\k},t)\ . \eqno(C11) \en By integrating eq.\ (C11), the
solution for the density fluctuations is

\eq {\tilde \delta }({\k},t)= -ik \int ^t _{t_{\rm rec}} dt' \, {{\tilde
v}_{\p}({\k} ,t') \over R(t')} \,. \eqno(C12) \en By multiplying by the
complex conjugate of eq.\ (C12), and then taking its average, we obtain

\eq\eqalignno{ \langle |{\tilde \delta}({\k},t) |^2 \rangle  &=k^2 \int ^t
_{t_{\rm rec}} {dt_1 \over R(t_1)} \,\int ^t _{t_{\rm rec}} {dt_2 \over
R(t_2)} \, \langle {\tilde v}_{\p}({\k},t_1){\tilde v}^*_{\p}({\k},t_2)
\rangle &\cr &=k^2 \int ^t _{t_{\rm rec}} {dt_1 \over R(t_1)} \,\sqrt
{\langle |{\tilde v}_{\p}({\k},t_1)|^2 \rangle} \int ^t _{t_{\rm rec}} {dt_2
\over R(t_2)}\,\sqrt{\langle|{\tilde v}_{\p}({\k},t_2)|^2 \rangle} &\cr
&=\biggl[k \int ^t _{t_{\rm rec}}{dt_1 \over R(t_1)}\sqrt{\langle |{\tilde
v}_{\p}({\k},t_1)|^2 \rangle} \biggr] ^2\,, &(C13)\cr} \en  where eq.\ (C10)
has been used; taking the square root of eq.\ (C13), we obtain

\eq \sqrt {\langle | {\tilde \delta}({\k},t)|^2 \rangle} = k \int ^t _{t_{\rm
rec}} {dt_1 \over R(t_1)} \sqrt {\langle |{\tilde v}_{\p}({\k},t_1)|^2}
\rangle\,. \eqno(C14) \en In other words,

\eq
 {\tilde \delta}_{\rm rms}(k,t) = k \int ^t _{t_{\rm rec}} {dt_1 \over
R(t_1)} \,{\tilde v}_{\rm rms}(k,t_1)\,. \eqno(C15) \en The differential
equation for ${\tilde \delta}_{\rm rms}$ follows via taking the time
derivative of eq.\ (C15),

\eq {\partial \over \partial t} \, {\tilde \delta}_{\rm rms} (k,t) = k
{{\tilde v}_{\rm rms} (k,t)\over R(t)} \,. \eqno(C16) \en

\bigskip \centerline{\bf Appendix D}  \medskip  In this Appendix, we show how
to derive eqs.\ (69) and (74). We begin with eq.\ (65),

\eq {\x \cdot {\dot{\v}}({\bf x},t_{\rm rec})} =\alpha \left\{\x
\cdot[(\x\times {\B}_{ff0}) \times {\b}_0] +\x \cdot[(\x\times {\b}_0) \times
{\B}_{ff0}]\right\} \,, \eqno(D1) \en where, again,  $\alpha \equiv 1/ 4 \pi
\rho_0 R^2_{\rm rec}$. By using the Fourier expressions for ${\v}$,
${\B}_{ff0}$, and ${\b}_0$ we obtain \eq i{\k}\cdot {\dot {\tilde
{\v}}}({\k},t_{\rm rec}) =\alpha\int d^3 {\k}_1\left\{
k^2[{\tilde{\b}}({\k}_1)\cdot {\tilde{\B}}_{ff}({\k}-{\k}_1)] - 
[({\k}+{\k}_1)\cdot  {\tilde{\B}}_{ff}({\k}-{\k}_1)][{\k}\cdot
{\tilde{\b}}({\k}_1)] \right\} \,, \eqno(D2) \en  where
${\tilde{\B}}_{ff}({\k})$ is the Fourier transform of ${\B}_{ff0}$ and
${\tilde{\b}}({\k})$ is the transform of ${\b}_0$. The ensemble averages of
both sides of eq.\ (D2) vanish. Again, we compute ${\tilde v}_{\rm
rms}(k,t_{\rm rec})$ by taking the square of both sides of eq.\ (D2) and then
ensemble average, with the following result: \eq\eqalignno{ \langle
|{\k}\cdot {\dot {\tilde {\v}}}({\k},t_{\rm rec})|^2 \rangle & = \alpha ^2 
\,{ V k^2 \over 2\pi^2} \,\int_{k_{\rm min}}^{k_{\rm max}}dk_1 k_1^2
\int^1_{-1}d{\mu}   \,{{\tilde b}^2(k_1)\,{\tilde B}^2_{ff}(|{\k}-{\k}_1|)
\over |{\k}-{\k}_1|^2 } &\cr & &\cr &\times \left\{ k^4 + k^2 k_1^2 (1 -
\mu^2 + \mu^4) - k^3 k_1 \mu (1 + \mu^2) \right\}. &(D3)\cr} \en  The
polynomial integrand in eq. (D3) is not the same as in eq. (A9) since
${\B}_{ff}$ and $b$ are assumed to be uncorrelated and separated in $k$-space
(scale separation). This assumption changes the non-trivial contributions in
eq. (A8). 

For the force-free  background magnetic field, we assume a  delta function
spectrum and write $$ {\tilde B}^2_{ff}(k) \equiv \langle B_{ff0}^2 \rangle
{\delta (k-k_0) \over {4 \pi k^2}}\,, \eqno(D4) $$ where we assumed that
$k_0$ is small compared to the wavenumber of the fluctuations about the
force-free background. Eq.\ (D3) now takes the form 

\eq\eqalignno{ \langle |{\k}\cdot {\dot {\tilde {\v}}}({k},t_{\rm rec})|^2
\rangle & =  {\alpha ^2 \langle B_{ff0}^2 \rangle \,V  \over 4 (4\pi)^3\,k
k_0^3} \int_{k_{\rm min}}^{k_{\rm max}}\,dk_1{{\tilde b}^2(k_1) \over k_1}\,
\biggl\{ k_1^8 -(4 k_0^2+2k^2) k_1^6 +(6k_0^4+2k^2 k_0^2) k_1^4  &\cr &
+(-4k_0^6+2 k^2 k_0^4 +16 k^4 k_0^2+2k^6)k_1^2 + k_0^8 -2k^2 k_0^6 +2k^6
k_0^2 -k^8 \biggr\}\,. &(D5)\cr} \en

For a delta function Ansatz for ${\tilde b}^2(k)$,  $${\tilde b}^2(k) =
\langle b_0^2 \rangle {\delta(k-k_{b}) \over 4 \pi k^2} \, ,$$  only modes
satisfying $k_b-k_0 \leq k \leq k_b+k_0$ are excited. With the use of eq.
(D5), the spectrum of the generated compressible modes is given by $$  k^2
{\dot {\tilde v}}_{{\p}\rm rms}^2 ({k},t_{\rm rec})  \simeq {4\; \alpha ^2 V 
k^{3}\over (4\pi)^4\, k_{\rm b} k_0}\,  \langle B_{ff0}^2 \rangle \langle
b_0^2 \rangle \, F_{\delta}(k,k_0,k_b)\, , \eqno(D6)$$ where
$$F_{\delta}(k,k_0,k_b) \equiv 
  1+{{ (k_b^2 - k_0^2)^4 - 2 k^2 (k_b^2 - k_0^2) (k_b^4 - k_0^4) +
 2 k^6 (k_b^2 + k_0^2) - k^8} \over 16 k^4 k_b^2 k_0^2}\,, \eqno(D7)$$
from which eq.\ (69) follows.
 
In the limit $k_b \to k_0$ and for $k \ll 2 k_b$, we recover the result in
eq. (A10)
 with an overall factor of 4 difference due to the  relation between 
$\langle B_{ff0}^2 \rangle$, $ \langle b_0^2 \rangle$, and   $\langle B_{0}^2
\rangle$.

 The power-law Ansatz can be written as  $${\tilde b}^2(k_1) = C k_1^q \, ,$$
for $k_{\rm min} \le k_1 \le k_{\rm max}$, and $\langle b_0^2 \rangle  = 4 \pi
C\;(k_{\rm max}^{q+3} - k_{\rm min}^{q+3})/(q+3) \simeq 4 \pi C\;k_{\rm
max}^{q+3}/(q+3)$ for $k_{\rm min} \ll k_{\rm max}$. Substituting this Ansatz
in  eq. (D5), the integration can be performed within the following range: 
for $k \geq  k_{\rm max}-k_0$,  the upper limit on $k_1$ is ${\rm max}(k_1)
=k_{\rm max}$, and  for $k \leq  k_{\rm max}-k_0$, 
 ${\rm max}(k_1) =k + k_0$; while for  $k \leq  \sqrt{k_0^2+k_{\rm min}^2}$,
the lower limit is  ${\rm min}(k_1) =k_{\rm min}$, and finally, for $k \geq 
\sqrt{k_0^2+k_{\rm min}^2}$, ${\rm min}(k_1) =\sqrt{k^2-  k_0^2}$. 
 The solution for specific choices of $k_0$, $k_{\rm min}$,  $k_{\rm max}$,
and $q$ is  plotted in  Figs. (2) and (3).  

Since $k_0 \ll  k_{\rm min} \ll  k_{\rm max}$, the most relevant  range of
the integration is $\sqrt{k_0^2+k_{\rm min}^2} \le k \le k_{\rm max} - k_0$.
In this range, the integral in eq.\ (D5) can be written as

\eq  k^2 {\dot {\tilde v}}_{{\p}\rm rms}^2 ({k},t_{\rm rec})  \simeq  
{\alpha^2 V \over 4 (4\pi)^4} \langle B_{ff0}^2 \rangle \langle b_0^2 \rangle
{  k_{\rm max}^5 \over \, k_0^3 k  } \,(q+3) \, F_q(k,k_0,k_{\rm max}) \, ,
\eqno(D8)$$ where  $$\eqalignno{ F_q(k,k_0,k_{\rm max}) \equiv & {1 \over
q+8} -  {4 \kappa_0^2 + 2 \kappa^2 \over {q+6}} + {6 \kappa_0^4 + 2
\kappa_0^2 \kappa^2 \over q+4} & \cr + &
 { 2 \kappa^6 - 4 \kappa_0^6 + 2 \kappa^2 \kappa_0^4 + 16 \kappa^4 \kappa_0^2
\over q + 2}  + {\kappa_0^8 - 2 \kappa^2 \kappa_0^6 + 2 \kappa^6 \kappa_0^2 -
\kappa^8 \over q} ] \,; &(D9)\cr} $$ 
$\kappa_0 \equiv k_0 / k_{\rm max}$ and  $\kappa \equiv k / k_{\rm max}$. In
the limit $\kappa_0 \ll \kappa$, $ F_q \simeq 16 \; \kappa_0^3 \;
\kappa^{q+5}\; $  and, therefore,  $$  k^2 {\dot {\tilde v}}_{{\p}\rm rms}^2
({k},t_{\rm rec})  \simeq   {\alpha^2 V \over 64 \pi^4} \; \langle B_{ff0}^2
\rangle \; \langle b_0^2 \rangle \; (q+3) \; k  \, \left({ k \over
k_{\rm max} }\right)^{q+3}  \,. \eqno(D10)$$ \bye